\documentclass[twocolumn]{aa}

\usepackage{graphicx}

\usepackage{txfonts}
\usepackage{subfigure}
\usepackage{orcidlink}
\usepackage{academicons}
\usepackage{tablefootnote}
\usepackage{gensymb}
\usepackage[normalem]{ulem}
\usepackage{placeins}

\makeatletter
\renewcommand*\aa@pageof{, page \thepage{} of \pageref*{LastPage}}

\begin{document}

   \title{The MIRI/MRS Library} 
   \subtitle{III. Detector-based optimal spectral extraction and high-contrast imaging applications}

   \author{Danny Gasman
          \inst{1}*\orcidlink{0000-0002-1257-7742}
          \and
          Ioannis Argyriou
          \inst{2}\orcidlink{0000-0003-2820-1077}
          \and
          David R. Law
          \inst{3}\orcidlink{0000-0002-9402-186X}
          \and
          Sandrine Juillard
          \inst{4,5}\orcidlink{0000-0001-6258-2673}
           \and
          Valentin Christiaens
          \inst{6,2,7}\orcidlink{0000-0002-0101-8814}
          \and
          Olivier Absil
          \inst{7} 
          \and
          Alistair Glasse
          \inst{8}\orcidlink{0000-0002-2041-2462}
          \and
          Karl D. Gordon
          \inst{3}\orcidlink{0000-0001-5340-6774}
          \and
          Patrick J. Kavanagh
          \inst{9}\orcidlink{0000-0001-6872-2358}
          \and
          Jane E. Morrison
          \inst{4}\orcidlink{0000-0002-9288-9235}
          \and
          Polychronis Patapis
          \inst{10}\orcidlink{0000-0001-8718-3732}
          \and
          G. C. Sloan
          \inst{3,11}\orcidlink{0000-0001-8718-3732}
          }

   \institute{Max-Planck-Institut f\"{u}r Astronomie (MPIA), K\"{o}nigstuhl 17, 69117 Heidelberg, Germany
   		\and
   		Institute of Astronomy, KU Leuven, Celestijnenlaan 200D, 3001 Leuven, Belgium
         \and
        Space Telescope Science Institute, 3700 San Martin Drive, Baltimore, MD, 21218, USA
        \and
         Steward Observatory, University of Arizona, Tucson, AZ 85721, USA
         \and
         Subaru Telescope, National Astronomical Observatory of Japan, National Institutes of Natural Sciences (NINS), 650 North A‘ohōkū Place, Hilo, HI 96720, United States
         \and
         Université Paris-Saclay, Université Paris Cité, CEA, CNRS, AIM, F-91191 Gif-sur-Yvette, France
         \and
         STAR Institute, Université de Liège, Allée du Six Août 19c, 4000 Liège, Belgium
        \and
        UK Astronomy Technology Centre, Royal Observatory, Blackford Hill Edinburgh, EH9 3HJ, Scotland, UK
        \and
        Department of Physics, Maynooth University, Maynooth, Co. Kildare, Ireland
         \and
        Institute of Particle Physics and Astrophysics, ETH Zürich, Wolfgang-Pauli-Str 27, 8049, Zürich, Switzerland
        \and
        Department of Physics and Astronomy, University of North Carolina, Chapel Hill, NC 27599-3255, USA
        \\
            *\email{dagasman@mpia.de}
             }

   \date{Received 25 March 2026 / Accepted 19 October 2026}

  \abstract
   {The James Webb Space Telescope (JWST) has now been operational for five years, and the associated data reduction pipelines have reached maturity. For the Mid-InfraRed Instrument (MIRI) Medium Resolution Spectrometer (MRS), one can generally get a good-quality spectrum by simply running the standard pipeline, especially for unresolved sources. However, users are starting to look deeper into the data, trying to detect the smallest features in the faintest objects, and find close companions.}
   {Following the lower level corrections we derived in Papers~I and II in this series, we now introduce a novel way to extract MIRI/MRS spectra: from the detectors directly. Currently the most common way to analyse data from integral field spectrometers, such as the MRS, is to reconstruct a spectral cube from the detector images. Theoretically, detector-extracted spectra could reach higher signal-to-noise ratios (S/Ns), and better handle bad pixels.}
   {We applied the charge migration and fringe corrections derived in Papers~I and II to generate stable and well-sampled point-spread function (PSF) models using the sub-pixel scan of calibration star 10~Lac observed in PID~3779 during Cycle 2. These PSFs were used in a detector-based optimal extraction algorithm building on past approaches, where the contributions per pixel were weighted by their noise properties and contribution to the total illumination. Furthermore, the PSF models were used to subtract the PSF of a bright central source from the detector, both by direct subtraction, and a principal component analysis (PCA) approach.}
   {We find that our optimal setup allows us to get a $>$1.5 times higher S/N on a faint source compared to the standard pipeline cube extraction. However, we approximately matched more involved cube processing, for which the extraction aperture is smaller and the background subtraction was performed more optimally. We were also able to subtract the PSF down to sub-percent levels, allowing us to extract spectra from faint companions. Currently, the high-contrast imaging extraction quality falls short of a cube-based extraction.}
   {The novel approach greatly improves on the current standard pipeline reduction, but it is only able to match the performance of more optimised cube processing.  We have not yet reached the fundamental noise limits. However, there are clear points of improvement that can be addressed using additional data. We demonstrate that detector-based processing of MIRI/MRS data has great potential for the characterisation of faint sources and close companions.}


    \titlerunning{The MIRI/MRS Library III}
    \authorrunning{D. Gasman et al.}
   \maketitle

%

\section{Introduction}
\label{sec:introduction}

The Medium Resolution Spectrometer (MRS, \citealt{ref:23ArGlLa}) of the Mid-InfraRed Instrument (MIRI, \citealt{ref:23WrRiGl}) on the James Webb Space Telescope (JWST, \citealt{ref:23RiPeMc}) is plagued by a variety of effects that limit its performance. One of the most well-known and impactful of these effects is detector fringing, where interference due to the reflectivity of the detector coatings creates a sinusoid pattern in the spectrum \citep[e.g.][]{ref:20ArWeGl}. A second lesser known, yet equally impactful, effect is charge migration, during which contrast between pixels causes photo-electrons to leak from a bright pixel to a less bright pixel \citep{ref:23ArLaRi}. The correction of the latter was addressed in \citet{paperI}, hereafter Paper~I, whereas the former was addressed in \citet{paperII}, hereafter Paper~II. The results from these two works in this series largely mitigate the aforementioned effects, which we subsequently show here to be essential to make detector-based spectral extraction and point-spread function (PSF) subtraction for the MRS possible. For more details regarding the specific mitigation methods, we refer the reader to these previous works.

\begin{figure}
\centering
	\includegraphics[width=\columnwidth]{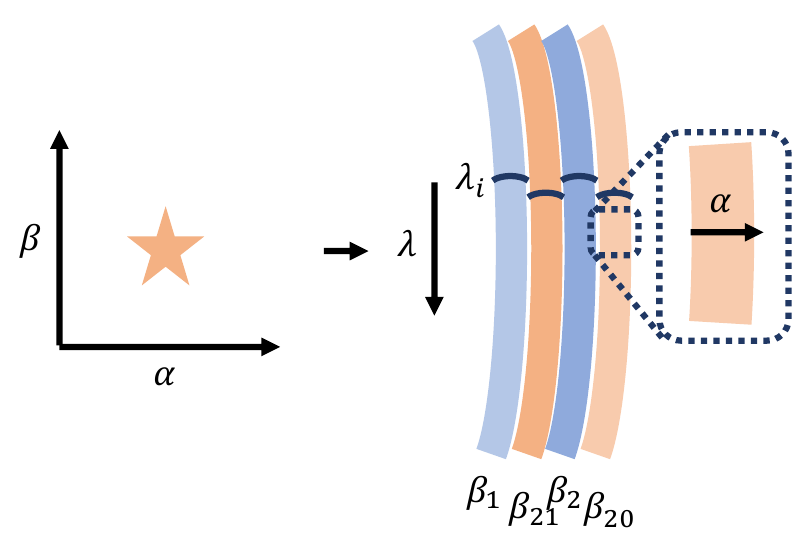}
	\caption{Schematic of the detector-based coordinate system.  We note that an $\alpha$ and $\beta$ are the spatial internal detector coordinates, while $\lambda$ indicates the spectral coordinates. The $\beta$ subscripts show the associated slice number, demonstrating how neighbouring on-sky $\beta$ coordinates are not directly next to each other on the detector.}
	\label{fig:data_schematic}
\end{figure}

\begin{figure*}
\centering
	\includegraphics[width=0.8\textwidth]{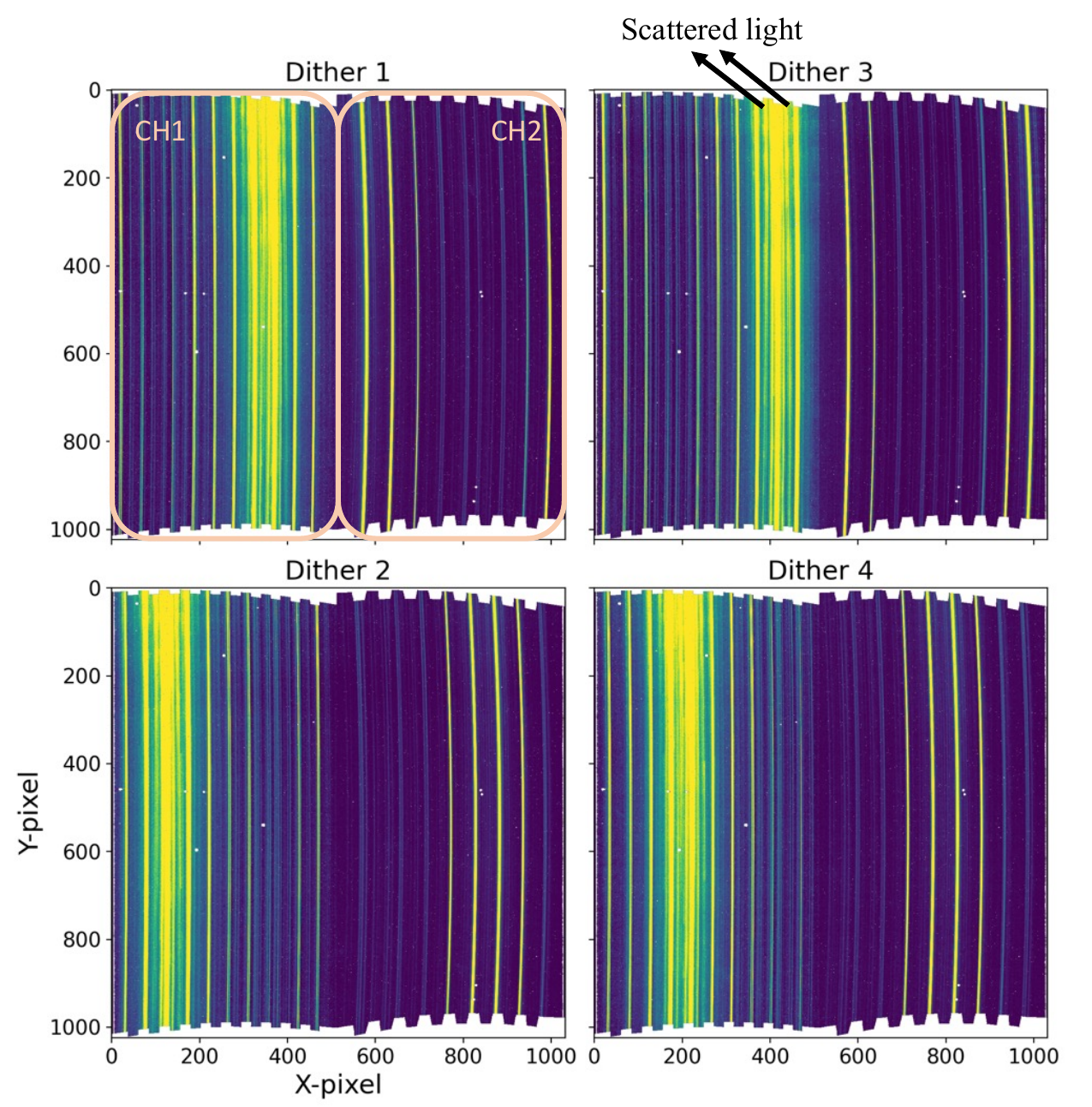}
	\caption{Demonstration for how an unresolved source appears on the detector for four different dither positions. Both channels 1 and 2 are projected onto the same detector. We can see the scattering light from the PSF in the interleaved slices around the slices that correspond to the on-sky coordinates that sample the PSF. We note that this is not a diffraction effect. These slices would be dark if there were no light scattering, as one can see on the channel 2 side where this effect is much smaller. Instead, internal scattering in the detector causes light to bleed into the neighbouring slices.}
	\label{fig:dither_schematic}
\end{figure*}

Currently, the analysis of MRS data relies on building spectral cubes. As discussed in detail in \citet{ref:23LaMoAr}, this involves a complex averaging of multiple pixels to create an evenly sampled datacube. While averaging multiple pixels should reduce spurious features, it does introduce correlated noise. Furthermore, while single bad pixels can be patched quite well by the current outlier detection algorithm, larger clusters of bad pixels still pose a significant problem and result in significant artefacts in the spectrum extracted from an aperture. Cubes do have the advantage of being straightforward to examine when a user is not well acquainted with the detector layout. However, when the data consist of a single unresolved source, or even a bright central source and a close companion, converting to an on-sky image is less important.

\begin{figure*}
\centering
	\includegraphics[width=0.8\textwidth]{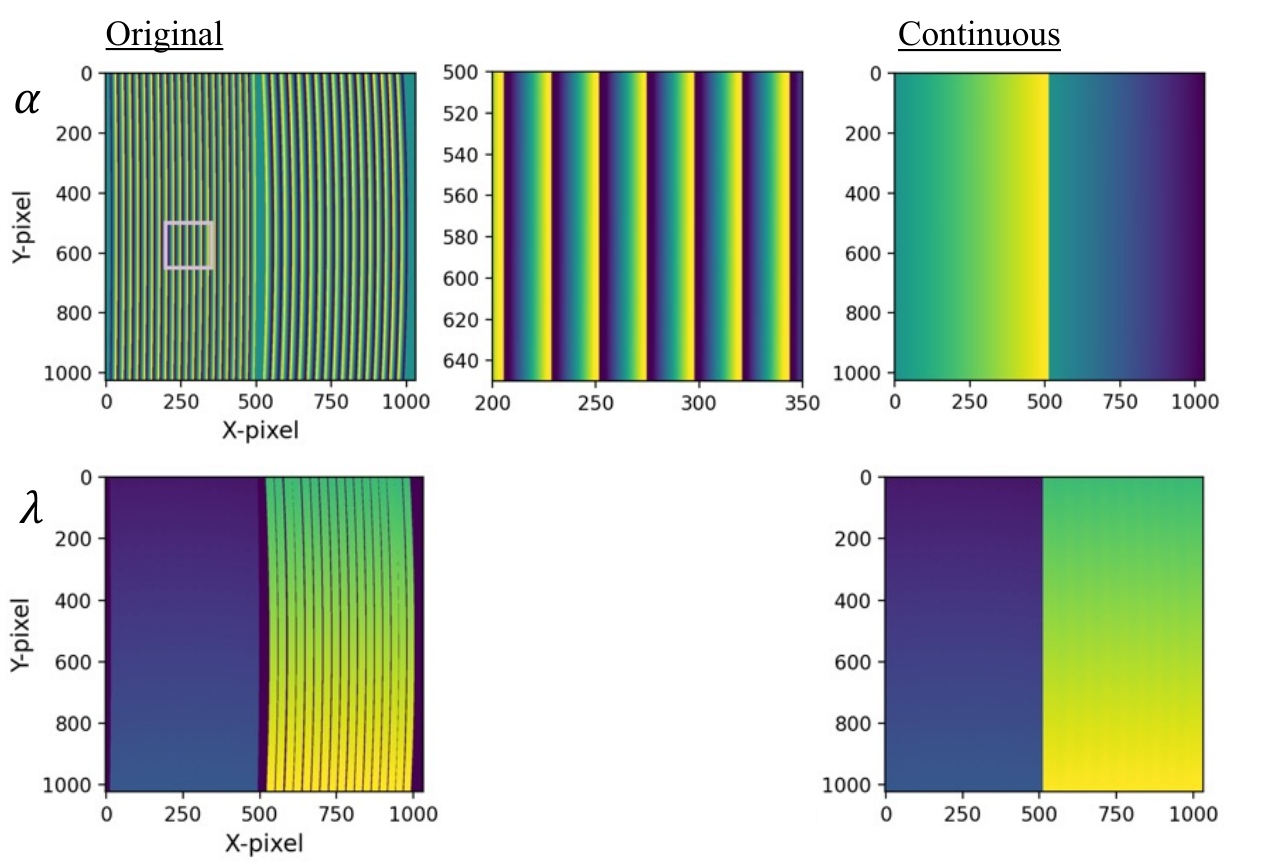}
	\caption{Continuous coordinate maps of $\alpha$ (top) and $\lambda$ (bottom). The left column shows the original map, the right column the continuous version. The centre panel at the top shows a zoomed in version of the original $\alpha$-map (the area is indicated by the square on the left) to demonstrate the changes of the coordinate within a slice.}
	\label{fig:cont_maps}
\end{figure*}

\begin{figure}[t]
	\centering
	\includegraphics[width=0.75\columnwidth]{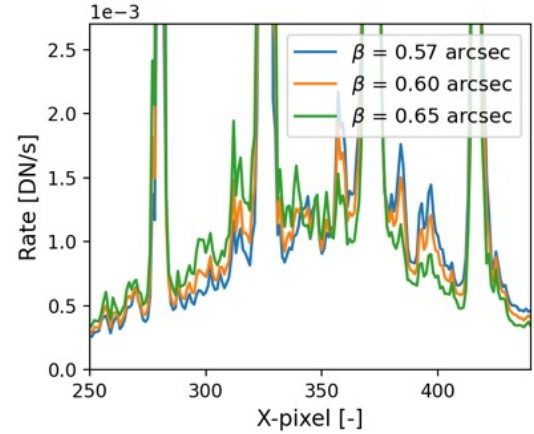}
	\caption{PSF models after combining the grid data (top). The models per $\beta$-pointing combined from three individual grid points are shown, zooming in on the change in scattered light with $\beta$.  The bottom two panels show how the models were combined per $\beta$ position (left), and how these three reference models are subsequently used to model the PSF at a different position within the grid (right).}
	\label{fig:psf_models}
\end{figure}

Instead of extracting spectra from a reconstructed cube, we aimed to extract spectra from the detector directly. Similar algorithms have been shown to result in an up to 30\% increase in signal-to-noise S/N in charge-coupled devices \citep[e.g.][]{ref:86Ho,ref:89Ma,ref:92Va}, and detector-based high-contrast imaging shows great promise for JWST's Near Infrared Spectrograph \citep[NIRSpec;][]{ref:24RuPeHo,ref:25MaRuMa,ref:26RuXuCh}. As a result of Papers~I and II, the MRS PSF is now much more consistent between observations, and the raster scan of 10~Lac from the calibration programme 3779 (PI: D. Gasman) provides the optimal dataset to generate PSF models for each nominal dither position. Furthermore, by working with the PSF on the detector, we can include scattered light into the model rather than removing it, recovering up to 20\% of lost light \citep{ref:23ArGlLa}. As a result, these models can not only be used to optimally extract spectra from the detector, but also to subtract a central bright source to uncover and extract the spectrum of a faint companion. We note that, as discussed in Papers~I and II, the calibration star only provides a sufficient S/N in the shorter wavelength range of the MRS (channels 1 and 2). We therefore set up the framework in the short wavelengths in this work, which is to be supplemented by suitable data in the future.

This paper presents the methods to generate the detector-based PSF models, how these can be used to extract spectra, and their high-contrast imaging applications. The required pre-processing steps and how the significantly distorted PSFs are fitted from the 3779 data are discussed in Sect.~\ref{sec:methods}. The results of the spectral extraction and high-contrast imaging can be found in Sect.~\ref{sec:results}. Finally, potential future improvements and a summary of the results are presented in Sects.~\ref{sec:discussion} and \ref{sec:conclusion}, respectively.

\begin{figure*}
\centering
	\includegraphics[width=\textwidth]{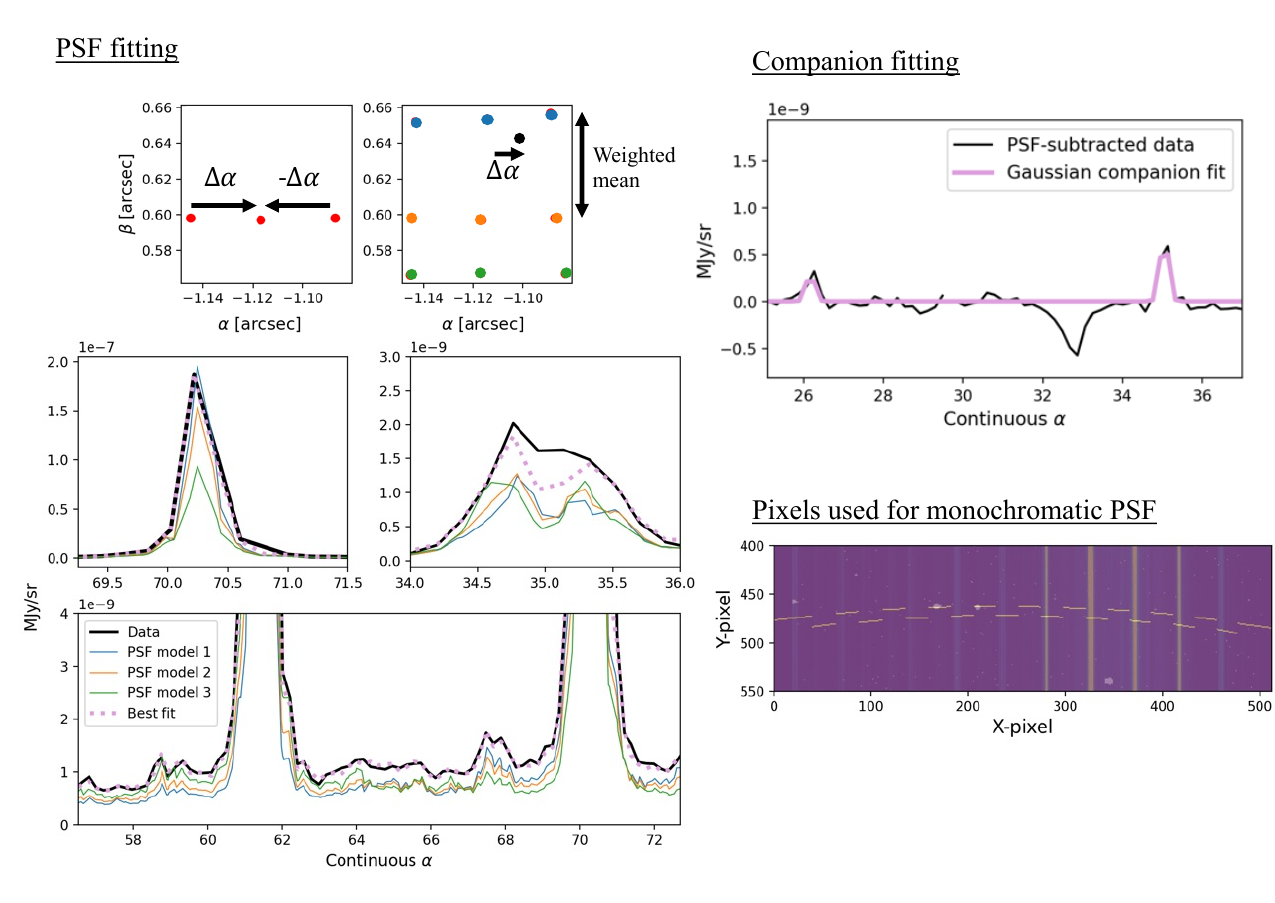}
	\caption{Demonstration of the PSF fitting process. On the left, we demonstrate the projection of the PSF models onto a calibration star by showing the details of a fit to a single wavelength bin. The grid points with the same $\beta$ are combined into one model (top left), yielding a single model per $\beta$ coordinate. The models are combined based on a weighted mean between $\beta$ coordinates and shifted in $\alpha$ until a best fit is found (bottom gallery). The top right shows an example of the per-slice gaussian fit of the companion. The number of slices to include can be set by the user, but for faint objects 1-2 slices is likely the best choice. Finally, the pixels used for the monochromatic PSFs shown here are highlighted in the bottom left corner. The yellow stripes overlaid on the original detector image constitute an iso-$\lambda$, or a single $\lambda$ coordinate.}
	\label{fig:fit_demo}
\end{figure*}

\section{Methods}
\label{sec:methods}

\subsection{The detector layout}
Prior to explaining the methods used in this work, we give an overview of the detector layout, in order to demonstrate the difficulties and differences from classical optimal extraction methods from slit spectrographs. Since the MRS is an integral field spectrometer, the detector effectively contains a set of slits that each represent a part of the sky image. The schematic in Fig.~\ref{fig:data_schematic} shows the along-slice ($\alpha$) and across-slice ($\beta$) coordinates,  representative of right-ascension and declination on-sky.  On the detector, the slices are interleaved, each with their own $\beta$ coordinate. Since they are interleaved, neighbouring slices on the detector are not neighbouring on-sky. The $\alpha$ coordinate then varies approximately horizontally within a slice, whereas the wavelength ($\lambda$) varies vertically within a slice. A single $\lambda$ shows up curved within a slice, and is shifted between slices, such that a single horizontal detector cut does not sample a single wavelength bin.

Due to the increasing PSF size with wavelength, the four different channels of the MRS are associated to a different image slicer with changing physical slice size, and a different number of slices.  Channel 1 has the most, with 21 slices, followed by 17, 16, and 12, for channels 2, 3, and 4, respectively. The MRS observes these channels simultaneously using two different detectors with a single channel covering either the left or right half, as shown in the top left of Fig.~\ref{fig:dither_schematic}.

The fact that the MRS is an integral field spectrometer means that the PSF-signature does not just cover a single slice, but practically all slices. Furthermore, since these slices are not neighbouring on the detector, parts of the PSF can be on opposite sides, which is visible for some dithers in Fig.~\ref{fig:dither_schematic}. An additional issue is that the scattering photons that cause fringing also cause a spatial scattering signature, as we show in Fig.~\ref{fig:dither_schematic}, where the area of strongest scattered light has been indicated using black arrows. This effect can be up to 20\% of the PSF, and bleeds into neighbouring slices that are not neighbouring on-sky. This scattering light is part of the PSF, and is best modelled as such. The final point to keep in mind is the fact that observations are dithered to limit the effects of clusters of bad pixels and properly sample the PSF, which is undersampled at short wavelengths, with less than 2 pixels per FWHM. While not visible in the rectified cube, where all dithers have been combined onto a single, centred PSF, in reality the dithers move across the entire detector. This is shown in the different panels of Fig.~\ref{fig:data_schematic}. The large distances mean that the PSF hits the detector at a different angle,  covers a slightly different pixel scale, and is affected by a different uncertainty in the coordinate map. The consequence of this is that it becomes non-trivial to extrapolate a PSF model from one location to another, at least when the $\beta$ offsets become large, shifting the centre of the PSF to a completely different set of slices. We can therefore only model the PSF accurately in the dither positions sampled by our reference observation.

\subsection{Detector-based PSF subtraction and optimal spectral extraction}
The process of subtracting the PSF and optimally extracting a spectrum on the detector using a PSF model as a weighting function is complementary. Both require a clean, well-sampled PSF model that can reproduce the data. Since the PSF and its scattered light have a specific shape depending on its location on the detector, the PSF was modelled based on the detector coordinate system, as we outline in more detail in Sect.~\ref{sec:psf}. 

This meant that we had interpolated models between different $\alpha$ and $\beta$ centroid coordinates, for all wavelength ($\lambda$) bins, based on the parameter space sampled by our grid. If the location of the science target on the detector was known, the model PSF could be produced at the same position, allowing for a small optimisation in coordinates and scale to accommodate for small uncertainties in the centroiding algorithm. This could be used to subtract the central bright PSF, or to extract a spectrum following the methods in \citet{ref:86Ho} and \citet{ref:89Ma}, which we discuss below. 

The main governing equation to extract a spectrum from the detector relies on three parameters: the normalised PSF model per pixel and wavelength bin ($P_{x\lambda}$); the background-subtracted, flux-calibrated signal ($S_{x\lambda}$); and the individual weights of the pixels ($W_{x\lambda}$). The optimal flux $f^\text{opt}_{\lambda}$, is then as follows:
\begin{equation}
	f_{\lambda}^\text{opt} = \frac{\sum_{x} W_{x\lambda} S_{x\lambda} / P_{x\lambda} }{\sum_{x} W_{x\lambda}} \text{.}
	\label{eq:opt_extr}
\end{equation}
Because of $W_{x\lambda}$, pixels that are relatively noisy, or do not contain a lot of flux from the PSF, automatically contribute less to the spectrum than pixels that sample bright parts of the PSF. This is due to its dependence on the variance of the science data ($V_{x\lambda}$) and $P_{x\lambda}$:
\begin{equation}
	 W_{x\lambda} = \frac{P^{2}_{x\lambda}}{V_{x\lambda}} \text{.}
\end{equation}
In order to be able to reject bad or noisy pixels while filling in the gaps, $P_{x\lambda}$ must be known at all pixel locations, and its sum per wavelength bin must be unity. In other words
\begin{equation}
	 \sum_{x}P_{x\lambda} = 1 \text{,}
	 \label{eq:psf_norm}
\end{equation}
such that the flux missing due to a cluster of bad or otherwise noisy pixels can always be known, removing spurious features.

\subsection{Pre-processing steps}
Prior to defining all of the parameters in Eq.~\ref{eq:opt_extr}, some pre-processing steps had to be applied. These were the standard pipeline steps described in Papers~I and II, where in this case no pipeline steps past selected \texttt{CALWEBB\_SPEC2} were applied, and the \texttt{straylight} step was skipped.

In order to include the scattered light, which is normally removed by the \texttt{straylight} step, the calibration and coordinate maps had to be extended to also include the areas between the slices. The maps for which this was done are the pixel flat, spectrophotometric calibration map, $\alpha$ coordinate map, and wavelength map. For the pixel flat and wavelength maps, the values from the slices are extrapolated to the inter-slice areas. The spectrophotometric calibration is simply a 1D calibration vector per wavelength \citep{ref:25LaArGo}, and was mapped to the extrapolated wavelength space on the detectors. Applying the pixel flat and spectrophotometric calibration map to the \texttt{\_rate} data resulting from \texttt{CALWEBB\_DETECTOR1} yielded the $S$-matrix in Eq.~\ref{eq:opt_extr}.

The $\alpha$ coordinate map was altered to be continuously increasing with each X-pixel on the detector, such that the PSF models could be consistently shifted across the entire detector-half in $\alpha$ space to match the science data. Therefore, the increments in $\alpha$ of the pixels are cumulatively added from X-pixel 0 to X-pixel 511 in case of channel 1, and from X-pixel 512 to X-pixel 1032 for channel 2. Examples of the continuous $\alpha$ and wavelength calibration maps can be found in Fig.~\ref{fig:cont_maps}. The continuous maps were applied to the science target to yield a flux-calibrated detector-based signal ($S$ in Eq.~\ref{eq:opt_extr}).

\begin{figure}[t]
	\centering
	\includegraphics[width=0.8\columnwidth]{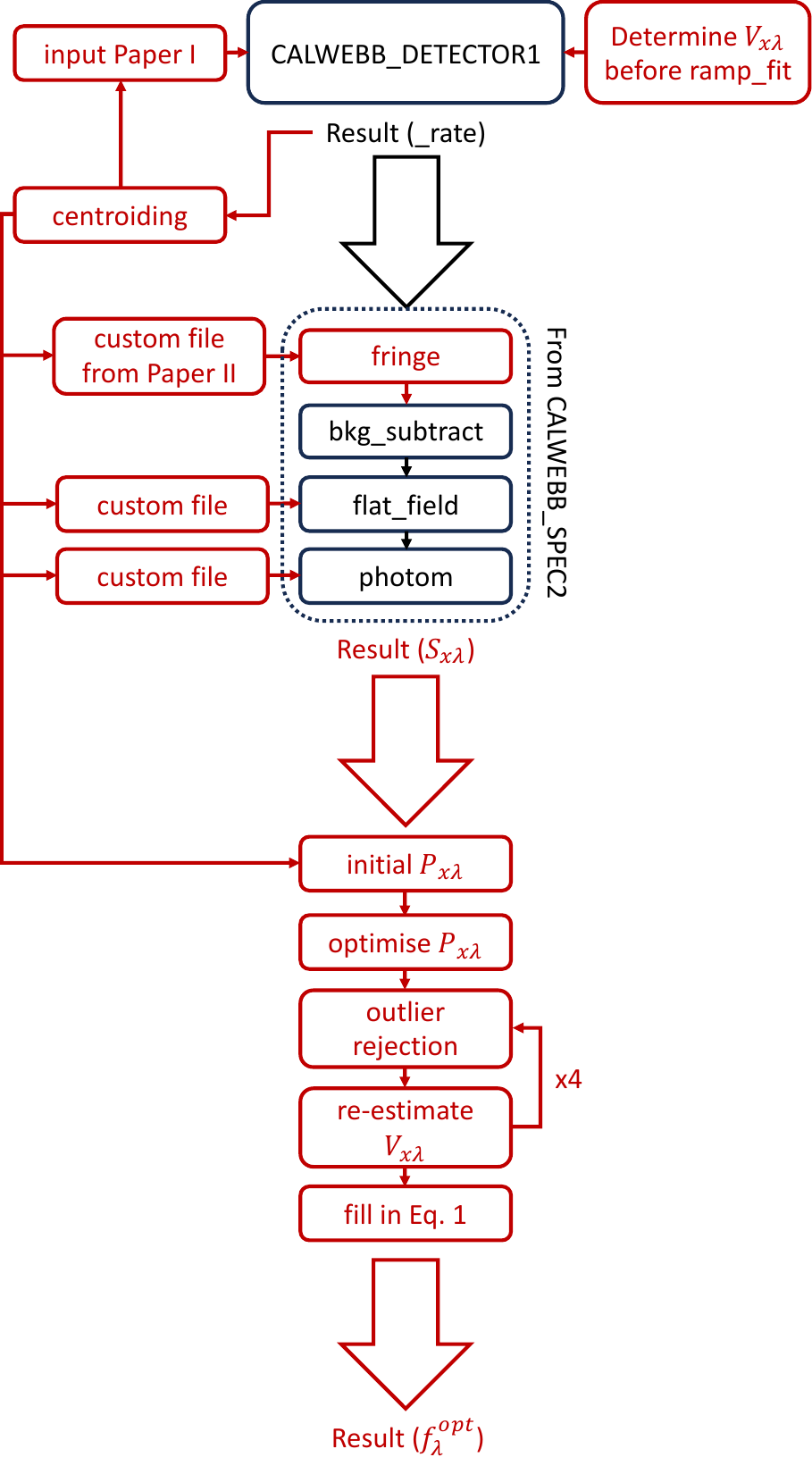}
	\caption{Flowchart of the optimal extraction pipeline in relation to standard pipeline steps. Red boxes indicate changes with respect to the standard pipeline steps or architecture.}
	\label{fig:pipe_schematic}
\end{figure}

\subsection{Generating the PSF models}
\label{sec:psf}
The next parameter in Eq.~\ref{eq:opt_extr} is the PSF model $P$. For this purpose, we used the sub-pixel scan of calibration star 10~Lac around the nominal dither positions from PID~3779.  Similarly to the case of \citet{ref:89Ma}, the MRS PSF is significantly distorted on the detector. On top of that, the MRS is an integral field spectrometer, resulting in a PSF that covers a large part of the detector space with the same spectral order covering multiple slices. While normally the MRS PSF is undersampled in the short wavelengths (channels 1 and 2), contrary to the case of \citet{ref:89Ma}, the sub-pixel scan provides a well-sampled reference dataset that can be used to generate PSF models, without making use of their sub-pixel sampling method.  To reproduce the complex distorted shape, we used iterative spline fits for every detector column, rather than single polynomials per column. This was done per set of three pointings that cover approximately the same $\beta$ coordinate to sample the same scattered light pattern,  the variation of which is shown in the top panel of Fig.~\ref{fig:psf_models}. For the same $\beta$, but different $\alpha$ the scattered light structure is the same. These three different samples were then recentred onto the same $\alpha$-centroid coordinate, as illustrated in the top left panel of Fig.~\ref{fig:fit_demo}.

Since we knew what the PSF looked like within the grid of coordinates, we could generate detector-based PSFs at the location of the science data. In order to do so, we ran a centroiding algorithm on the detector to get a first estimate for the location of the PSF. The two $\beta$-pointings of our grid closest to the science data were then shifted to the retrieved $\alpha$-centroid, and combined using a weighted mean based on the distance of the science point to the grid, as illustrated in the bottom right panel in Fig.~\ref{fig:fit_demo}. Then, by minimising the residuals per wavelength bin, we allowed the $\alpha$ and $\beta$ centroid and PSF amplitude to change slightly to optimise the placement of the PSF. An example of the result can be found in the left column of Fig.~\ref{fig:fit_demo}, which also provides a good reference of comparison for the output model and input data of the user. For PSF subtraction, this optimisation and scaling was kept per individual wavelength bin. For optimal extraction this introduced additional noise, therefore we smoothed the updated PSF coordinates as a function of wavelength and used these to generate the PSFs in each wavelength bin. The scaling factor was of no consequence here, since $P$ had to be normalised per wavelength bin (see Eq.~\ref{eq:psf_norm}).

\subsection{The weight map}
The last missing link in Eq.~\ref{eq:opt_extr} is the weight map $W$, which depends on both $P$ (which we derived above) and the variance map $V$. The methods in \citet{ref:86Ho} and \citet{ref:89Ma} relied on iterative measurements of the pixel variance while fitting the PSF model. After running the \texttt{CALWEBB\_DETECTOR1} stage, the \texttt{\_rate} files contained variances for the read and Poisson noise of the observation. Therefore, we could simply take these variances and move forward. However, since we have the raw ramps available to us, an alternative would be to measure the variance of the samples around the linearised ramps. After testing both methods, we used the variance measured from the raw data itself, as it resulted in a significant improvement in S/N compared to the current pipeline output.

The variance map defined above presented a first estimate ($V_0$). After placing the PSF model onto the detector as described above, we looped through a sequence of recalculating the variance and rejecting outliers similarly to \citet{ref:86Ho}. The variances were then updated each time as follows:
\begin{equation}
	V_{x\lambda} = V_0 + |f_{\lambda} P_{x\lambda}| / C_{x\lambda} \text{,}
\end{equation}
where $f_{\lambda}$ is the optimal extracted flux in a wavelength bin, and $C$ the spectrophotometric calibration map. We divided by $C$ since our variance estimates were in data numbers (DN), meaning the variance update had to be in the same units. While our initial $V$ already presented a good estimate, performing this loop resulted in small, but significant improvements in the spectra.

\subsection{Outlier rejection}
An advantage of knowing exactly what $P_{x\lambda}$ is meant to be for each pixel in an observation, is that outliers can easily be rejected. As mentioned above, after fitting the PSF onto the data, we updated the variance and rejected outliers in a loop. A pixel was considered an outlier when
\begin{equation}
	(S_{x\lambda} - f_\lambda  P_{x\lambda})^2/V_{x\lambda} > 5\sigma \text{,}
\end{equation}
where $\sigma$ is the standard deviation over the same quantity of the entire wavelength bin:
\begin{equation}
	(S_{\lambda} - f_\lambda  P_{\lambda})^2/V_{\lambda} \text{.}
\end{equation}
The full spectral extraction process with respect to the current standard pipeline steps is illustrated in Fig.~\ref{fig:pipe_schematic}.

\subsection{Background subtraction}
While the shortest wavelengths of MIRI still experience limited thermal background due to the low operating temperatures, it already presented problems in channel 2. For the optimal extraction applications presented here, we dealt with the background in our test cases in two ways: (1) use the test observation's dedicated background if available; and if not, (2) perform pair-wise dither subtraction if the object is faint. We note that, as already discussed in Paper~II, contrary to the current pipeline procedure, we subtracted the background after fringe removal, to keep the spectral baseline at the correct level.

\subsection{Updates with respect to Paper II}
While testing variations of methods and reference files to find the cleanest possible outcome, we noted that making sure all maps and reference files were as noise-free as possible made a clear difference in the S/N of the spectrum. This also applied to the fringe flats from Paper~II. In that work, we already discussed the difficulty of extracting a fringe pattern that is as representative as possible while not including other systematics. However, here we applied a filter to the fringe flats to suppress some of the high-frequency noise. Too strong a filter resulted in large fringe residuals, whereas no filter resulted in more noise in the extracted spectra.  We opted for a 1D gaussian filter, assuming $\sigma = 2$ and a  Savitzky-Golay filter with a window length of 3 and a polynomial order of 2. A more optimal method for removing the noise is left for future updates.

\subsection{Principal component analysis}
\label{sec:pca}
In addition to a direct approach for PSF subtraction using the models, we also experimented with a principal component analysis (PCA, see e.g. \citealt{ref:07JeBlSi}, and \citealt{ref:12SoLaJa}) approach with the grid as a reference. In this case, we used each individual dataset, separated by band and dither position, as input. Since the SEDs of objects to remove can vary widely depending on the science case, the PSF is scaled and subtracted per wavelength bin.

For the PCA models, we then experimented with either applying PCA per wavelength bin (resulting in a maximum of 9 components), or expanding the matrix by assuming each detector row is a separate dataset, allowing us to extract over 500 components (512 for channel 1, 520 for channel 2). This information was then used in a singular value decomposition (SVD):
\begin{equation}
	\text{\textbf{X}} = \mathbf{U\Sigma E}^{\text{T}} \text{,}
\end{equation}
where \textbf{X} the matrix containing the flattened vectors of each individual dataset, \textbf{U} is the unitary matrix, \textbf{$\Sigma$} the vector of singular values, and \textbf{E} the matrix containing the principal components. We used \texttt{numpy.linalg.svd} to solve for the individual matrices, where the mean is not subtracted from the raw input data \textbf{X}. If $n$ is the number of samples and $m$ the number of features (i.e. pixels), the matrices have dimensions ($n$,$m$), ($n$,$n$), ($n$), and ($n$,$m$), respectively.

The PCA aims to identify the most stable features in the data, where lower ranked PCs contain contributions of increasingly smaller variance.  These features can then be projected onto the science image to reproduce the PSF, and subtract it. For a PSF that is expected to contain multiple important contributing features that may vary from observation to observation,  a simple mean (only keeping the first PC) is not be sufficient, and a larger set of references may be required to properly reproduce science data. We discuss the best-performing methods in Sect.~\ref{sec:subtraction}

\subsection{Extracting a companion outside the grid}
The companion used as a test case here lies about $0\farcs7$ from its primary, which is well outside our 60$\times$60~mas grid. Extrapolating the PSF models far outside the grid boundaries could be done, but did not yield the accuracy needed to get a high-fidelity spectral extraction. Furthermore, the bright primary did land in the nominal dither positions, and was therefore a great candidate for PSF subtraction using the grid. Extracting the companion then required a slightly different treatment compared to the method defined above.

The signal $S$ in Eq.~\ref{eq:opt_extr} must be background subtracted, but the dataset used to test the high-contrast imaging applications in this work did not have a dedicated background. We tried to subtract the background prior to further processing using PCs extracted from around 100 background observations and masking the central PSF. However, we found that the PSF covers such a large part of the detector, especially for bright sources, that we could not mask enough of the PSF and get a good background estimate. Instead, generating PSF PCs based on reference data that still included the background worked better. We discuss other potential improvements in Sect.~\ref{sec:discussion}.

Since the companion was very faint, a complex PSF model as derived above was not necessary to extract a spectrum; most of the details were buried in the noise. Instead, we performed simple gaussian fits to the brightest one or two IFU slices, depending on how clearly the signal was present in the fainter slice (two for channel 1, one for channel 2). Here we only fit centroids and amplitudes, and assumed the FWHM per wavelength resulting from Paper~I. The centroids were then fit with a second order polynomial to yield a smooth and noise-free solution. The resulting fit compared to the data is shown in the right column of Fig.~\ref{fig:fit_demo}.  Using our grid PSF model, we then estimated how much flux was missing when only including the PSF peaks from some slices to apply as a correction in the optimal extraction routine.

To further limit the noise contributions from this noisy data-driven approach, we applied an aperture on these gaussians when extracting the fluxes. We used an aperture of 1.2$\times$FWHM in the $\alpha$-direction, since the $\beta$-direction was already limited by the number of slices we included. To make up for the resulting flux losses, an aperture correction was applied in the optimal extraction routine.

\subsection{Contrast curves}
In order to provide a comparison that is more familiar to those working on high-contrast imaging applications, we generated contrast curves. A typical method is to inject companions into a PSF-subtracted dataset, and find the lowest flux for which a 5$\sigma$ detection can be achieved. This companion is then injected at smaller and smaller radial distances at different azimuths. Our detector-based methods make this less straightforward, so we outline the adjustments to this method here.

Firstly, the spatial on-sky offsets are mapped in a non-trivial way onto the detector, as we presented in Fig.~\ref{fig:data_schematic}. Due to the image being sliced into different $\beta$ coordinates, the optimal observing programme sets the position angle of the planet to maximise its offset in $\alpha$ instead. This would maximise the pixel-based separation on the detector, and prevent the spectral streaks of the primary and secondary from directly overlapping. We therefore limited the explored parameter space to $\alpha$ offsets.

Due to the less than critically sampled PSF on the MIRI/MRS detectors, with less than two pixels sampling one FWHM of the PSF, for faint objects the detectable flux is predominantly present in a single pixel.  Most of the rest of the flux is buried in the noise instead. Therefore, we injected the model PSF, found the brightest pixel per wavelength bin, and calculated the flux required to reduce it to the 5$\sigma$-level. By measuring the fraction between the primary's flux, and the retrieved secondary's flux, we calculated the contrast between the two.

We then generated the curves for two different sets. Firstly, we explored the optimal achievable contrasts for a read and photon noise-limited observation. This required us to set up a detector-based exposure time calculator. We first simulated the image, by injecting a primary star at the expected dither positions using our detector-based PSF models. For the channel 2 simulations, where the background becomes more prominent, we added a constant flux of 0.142~DN/s and 0.168~DN/s for sub-bands A and B, respectively. These values are based on the median of around 100 background observations, though we note that the MIRI detector background behaviour is known to be quite variable and erratic, even when not considering zodiacal light. We did not simulate band 2C, due to reduced quality of the reference files. Based on this flux distribution per pixel, we built ramps per pixel using the spectrophotometric calibration map. For each frame, we used the flux to generate the next expected frame. To this frame, we added read and photon noise based on the known read noise performance for the former, and a Poisson distribution for the latter. We do this for a predefined number of frames and integrations (where for each integration we simply simulate another ramp), in the case of this work 50 frames and 4 integrations, amounting to a total time of 37 minutes of observing time when doing four dithers. Then, similarly to what the JWST pipeline does, we built the calibrated detector image by fitting these ramps. Since we did not simulate the non-linearity on the ramps, and rather assumed this to be perfect, directly performing a simple linear fit was sufficient for this purpose. We simulated the four dithers for a central star ($\sim$1~Jy at 5~$\mu$m) in this way.

For the practical case, which does not solely include read and photon noise, we used the subtracted detector image of the bright test case $\delta$~Umi. This represents the non-ideal case, and demonstrates that there is still room for improvement in the presented methods. For all cases, we present the mean of these curves for all wavelength bins per band.

\section{Results}
\label{sec:results}

\begin{table}[h]
\caption{Summary of test data used in this work.}
\centering
\label{tab:test_obj}
\begin{tabular}{llll} \hline
\textbf{Source}     & \textbf{Type}    & \textbf{PID}  & \textbf{PI}          \\ \hline \hline
$\delta$~Umi & Calibration star & 1536 & K. Gordon   \\
GY92 141   &  Faint YSO     & 3886 & S. Grant    \\
GQ~Lup   & YSO+companion    & 1640 & A. Banzatti \\
\hline
\end{tabular}
\end{table}

We use three datasets for the tests below: one calibration star (the bright case for PSF subtraction tests), one faint young stellar object (YSO, the faint case), and one object known to have a faint companion (high-contrast imaging test-case). The objects and their programme IDs are summarised in Table~\ref{tab:test_obj}.

\subsection{Optimal extraction of a single point source}

\begin{table}[h]
\caption{S/N measured per object, extraction method, and band.}
\centering
\label{tab:snr_perf}
\resizebox{\columnwidth}{!}{
\begin{tabular}{cccccc} \hline
     & \multicolumn{3}{c}{GY92 141}  & \multicolumn{2}{c}{GQ~Lup B}  \\
Band &   Cube    &   Cube   &   Detector  &  Cube    & Detector     \\
  & (2$\times$FWHM)  & (1.5$\times$FWHM) &  & & \\ \hline \hline
  1A   &    17.5    &   28.3     &    25.9            &        16.1    &   10.3  \\
   1B  &      16.2    &    30.5    &   30.9            &      17.9      &  9.2  \\
 1C    &      19.2    &    41.6    &    39.0          &      14.9      &  9.9   \\
  2A   &    22.2    &   48.7     &    37.2             &     12.5       &   3.3   \\
 2B    &     17.4    &    40.5    &   30.3         &     16.4       &  3.3  \\
 2C	&	20.0	&	49.5	&  38.2 	&	8.7	&  1.9  \\ \hline
\end{tabular}
}
\end{table}

\begin{figure}[t]
	\centering
	\includegraphics[width=\columnwidth]{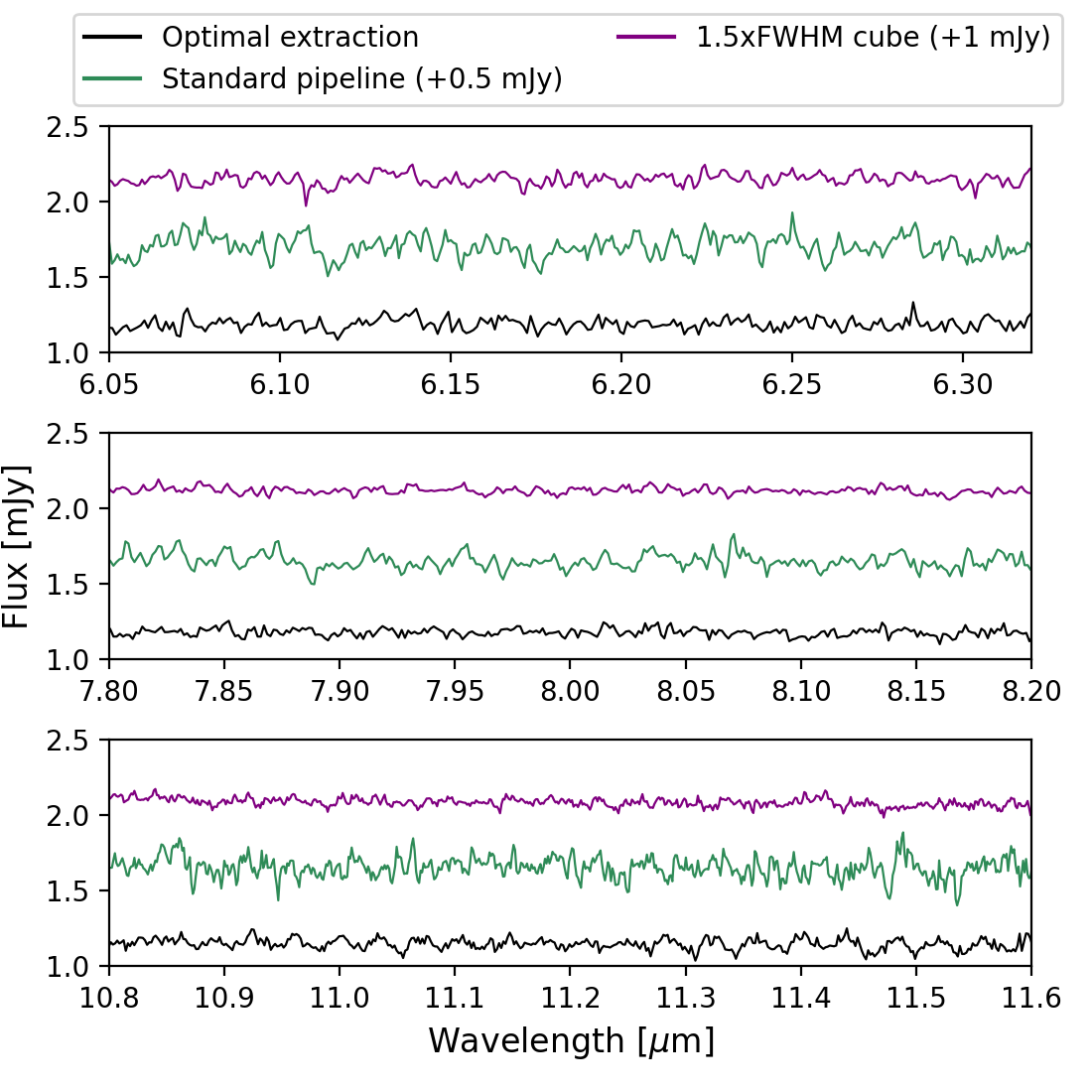}
	\caption{Extracted spectra of a faint source (GY92 141) zoomed on different wavelength regions,  using several extraction methods. The black line shows the detector-extracted spectrum, the green line the current standard pipeline output (which automatically assumes an aperture of 2$\times$FWHM), and a more optimised cube-based extraction in purple (courtesy of S.  Grant, assuming an aperture of 1.5$\times$FWHM).}
	\label{fig:disk_spec}
\end{figure}

First, we tested the detector-based, optimal extraction approach. For this purpose, we applied the methods to GY92~141, which is a single, faint, and unresolved object. This is where optimal extraction is expected to provide the largest improvements. We show details of the spectra in Fig.~\ref{fig:disk_spec}. When running the current standard pipeline without changing any parameters, the standard point source extraction aperture is 2$\times$FWHM, and the background is subtracted in an annulus. We clearly see a significant improvement that the detector-based extraction approach provides compared to the cube-based approach using this larger aperture. When measuring the S/N, summarised in Table~\ref{tab:snr_perf}, we note a $>$1.5 times improvement in this case. However, a more optimal cube-based extraction setup for faint sources involves a 1.5$\times$FWHM aperture (any smaller and PSF-sampling artefacts tend to become a problem; \citealt{ref:23LaMoAr}), and a pair-wise dither subtraction to remove the background. When using that approach, the S/N slightly exceeds that of the detector-based extraction. Looking at the top panel in Fig.~\ref{fig:disk_spec}, we do see that in some parts bad pixels are handled a bit better in our method, though this is subtle. Finally, the pair-wise background subtraction can result in slightly lower flux in the cube extraction case, due to negative lobes from the PSF of the other dithers. This is mitigated in the optimal extraction case, where negatives are removed and replaced with the expected PSF, much like bad pixels.

In the bottom panel, we note that the main culprit for the larger discrepancy in S/N in channel 2 is in fact residual fringing, which causes a distinct residual periodicity in the optimal extraction that is not present in the cube extraction. This is not a signature present in the source itself, but rather an instrumental effect. This is in line with our results in Paper~II, where we find that using the pointing-based defringing approach is trickier in channel 2. Future improvements in this regard should result in a closer match in S/N between the two spectra.

\begin{figure*}[h!]
	\centering
	\includegraphics[width=0.9\textwidth]{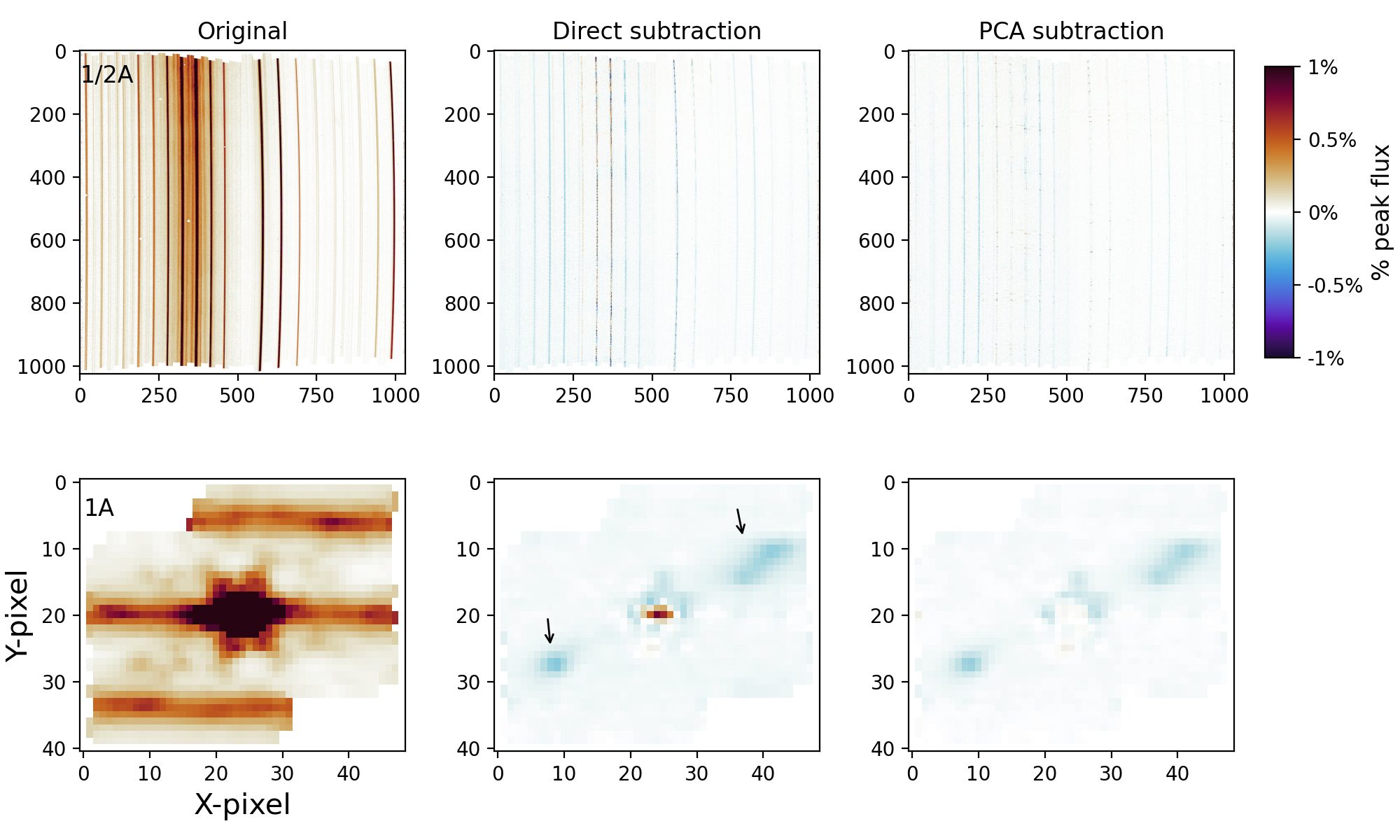}
	\caption{Performance of the different detector-based PSF subtraction methods, applied to a single calibration star ($\delta$~Umi).  The first row demonstrates the detector-based results for the different methods on sub-band A for a single dither, whereas the second row shows the results in the cube for band 1A. The first column presents the original data, the second column the residuals of direct subtraction using the detector-based models, and the third column the subtraction using PCA. The oversubtraction due to persistence is highlighted by the black arrows. We note that the cube panels present the median-collapsed cubes, in percent of the peak of the original data, whereas the detector-based results are a percent of the maximum flux in the image. Colourmap from \citet{ref:cmasher}.}
	\label{fig:star_subtr}
\end{figure*}

\subsection{Residuals on PSF subtraction}
\label{sec:subtraction}

Next, we test the performance of detector-based PSF subtraction applied to both the calibration star $\delta$~Umi and GQ~Lup, the latter of which hosts a companion \citep{ref:05NeGuWu}.  We show a single sub-band of both the subtraction on the detector for a single dither and the dither-combined cube image for illustration purposes. Residuals for the cubes of other bands can be found in App.~\ref{app:psf_sub}. The results from both the direct and PCA subtraction methods, applied to both test cases, can be found in Figs.~\ref{fig:star_subtr} and \ref{fig:gqlup_subtr}. The left column shows the original data, including the scattered light artefacts, the central column the direct subtraction, and the last the PCA subtraction. $\delta$~Umi benefits from a dedicated background, which we use in this case to remove the background prior to PSF subtraction. We discuss the individual subtraction methods in more detail below.

\subsubsection{Direct subtraction}
For direct subtraction, the PSF models are placed on the source in the detector space per wavelength bin, similarly to how the PSFs are defined for the spectral extraction. This time, though, the signal is subtracted. As seen in Fig.~\ref{fig:star_subtr}, in the calibration star test case we generally reach percent-level residuals in the median-collapsed cube image. Considering $\delta$~Umi and the reference star were observed with a different number of integrations and frames in completely different observing Cycles, this performance is quite good.  We note that the negative residuals in the lower left and upper right of the cube images are due to persistence in the detectors from previous dithers \citep[see also Sect. 3.4 in][]{ref:23ArGlLa}. Since our reference was observed such that it significantly saturates the detectors in the shorter wavelengths, the persistence signature included in the models is stronger and results in oversubtraction.

If we instead move on to our test case with a companion in Fig.~\ref{fig:gqlup_subtr}, the central source again shows percent-level residuals, and the companion, south of the primary, clearly shows up in all bands. Similarly to $\delta$~Umi, the observing setup was not consistent between the reference and test data, demonstrating the power of our corrections from Papers~I and II in getting a consistent PSF.  We note that as we move to longer wavelengths, the background starts to show a slight oversubtraction. This is due to the fact that the background was not subtracted in the test data prior to PSF subtraction, due to a lack of dedicated backgrounds.

\subsubsection{Subtraction using PCA models}

\begin{figure*}[h!]
	\centering
	\includegraphics[width=0.9\textwidth]{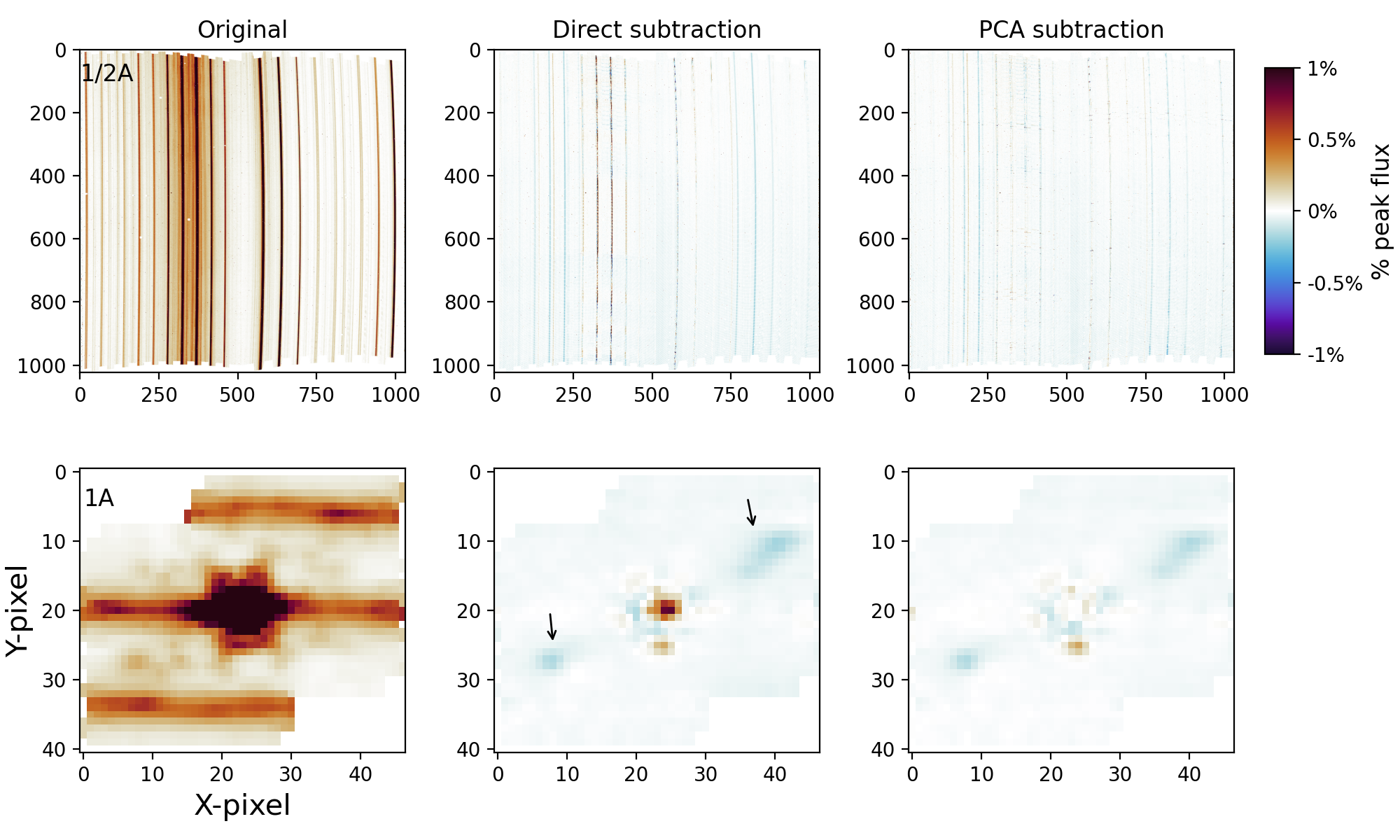}
	\caption{Same as Fig.~\ref{fig:star_subtr}, applied to an object known to host a companion (GQ~Lup A\&B) at a separation of $\sim0\farcs7$.  We indicate the companion with a circle, which is visible at the percent level below the central PSF.}
	\label{fig:gqlup_subtr}
\end{figure*}

The right-most columns in Figs.~\ref{fig:star_subtr} and \ref{fig:gqlup_subtr} present the PCA subtraction results. Here, we are able to improve the residuals even further. Both the PSFs of the calibration star and that of the central object hosting a companion are subtracted to well below 1\%. The residuals in the core are even $<$0.02\%. In the case of Fig.~\ref{fig:gqlup_subtr}, we now include the background in the PCA as well, which allows us to do a slightly better job at subtracting the PSF in channel 2. It can also be seen that fewer negative residuals in the background remain compared to the direct subtraction in this case, since the PSF itself is not used to compensate for the lack of background subtraction.

Our PCA approach can be set to be more or less aggressive. As discussed above, we can either use our more conservative approach, where we only generate PCs per wavelength bin, or we combine all wavelength bins into one to evaluate the many detailed ways in which the PSF can change. The former is the better and more stable approach, resulting in fewer systematics in the subtraction. Furthermore, it limits the variations to changes within the observations, rather than also with wavelength. We also note that some dithers appear to be more easily removed than others, which seems to be connected to whether or not the signal lands on a more distorted part of the detector. While the effect is limited when the goal is simply to detect the companion, as we already clearly do in Fig.~\ref{fig:gqlup_subtr}, incorrect subtraction results in clear systematics in the subsequently extracted spectrum, as we show next. We suspect that supplementing our data by more unresolved sources would allow us to overcome this hurdle.

\subsection{The spectrum of a faint companion}

\begin{figure}[t]
	\centering
	\includegraphics[width=\columnwidth]{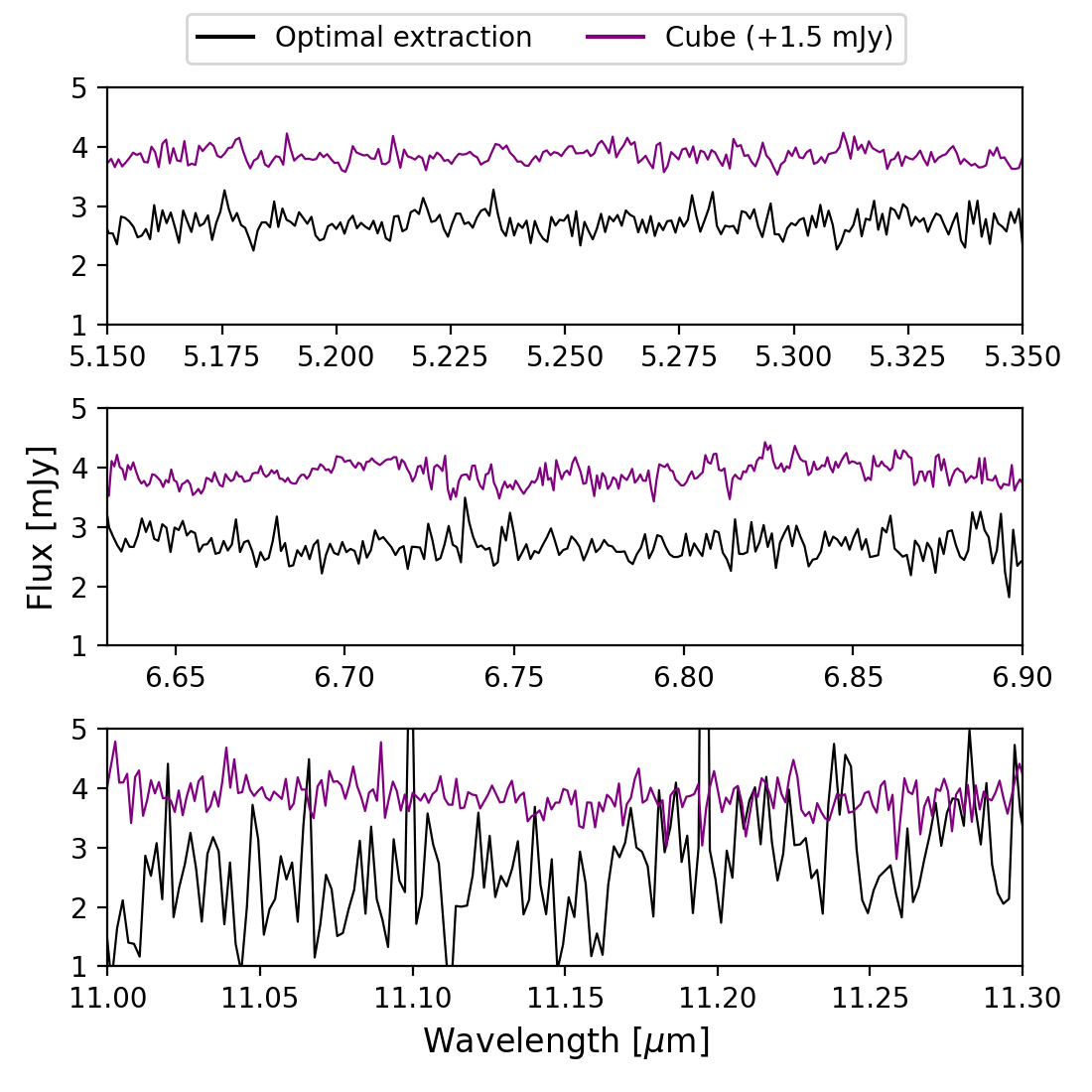}
	\caption{Extracted spectra of a faint companion (GQ~Lup B) zoomed on different wavelength regions,  using two extraction methods. The black line shows the detector-extracted spectrum, and the currently best cube-based extraction (courtesy of G. Cugno). The quality of the detector-based extraction degrades significantly towards 11~$\mu$m, due to the lower S/N on the reference files derived from the calibration star. This will be mitigated in the future using similar data on a red source.}
	\label{fig:companion_spec}
	\vspace{-1.5mm}
\end{figure}

After subtraction, we extract a spectrum from the companion. The resulting spectrum is shown in Fig.~\ref{fig:companion_spec}, along with the cube-based extraction from \citet{ref:24CuPaBa}. The comparison in measured S/N is again presented in Table~\ref{tab:snr_perf}.

The top two panels in Fig.~\ref{fig:companion_spec} demonstrate that, in channel 1 where the background contribution is still limited, we reach quite good quality spectra. Furthermore, in the second panel, the cube extraction shows some extra oscillation due to the PSF sampling, which is less prevalent in the detector-extracted spectrum. Nevertheless, we still do not reach the same S/N. This becomes especially problematic in channel 2 (bottom panel), where the systematics and residual noise become clearly worse than the cube extraction.

The reason for the clearly worse performance in channel 2 is likely the background contribution. Removing the PSF from the detector plane has proven to be quite challenging in general, and the addition of a residual background does not make this easier. As mentioned above, this results in significant systematics in specific dither positions, generally the second and the fourth, with the fourth being worst overall. This means that in the channel 2 spectrum shown, we have only combined two dithers, which already means we cannot reach the full theoretical S/N the total observing time would imply. 

Experimenting with methods to perform the PCA and subtraction demonstrates that this is highly sensitive to the amount of information available.  We discuss this in more detail below. We believe that adding a few more PSFs to the PCA and a better method to deal with the background would help us overcome this issue, and reach a similar performance to that of channel 1. Further improvements to the method are discussed below.

\subsection{Contrast curves}

\begin{figure*}[t]
	\sidecaption
	\includegraphics[width=12cm]{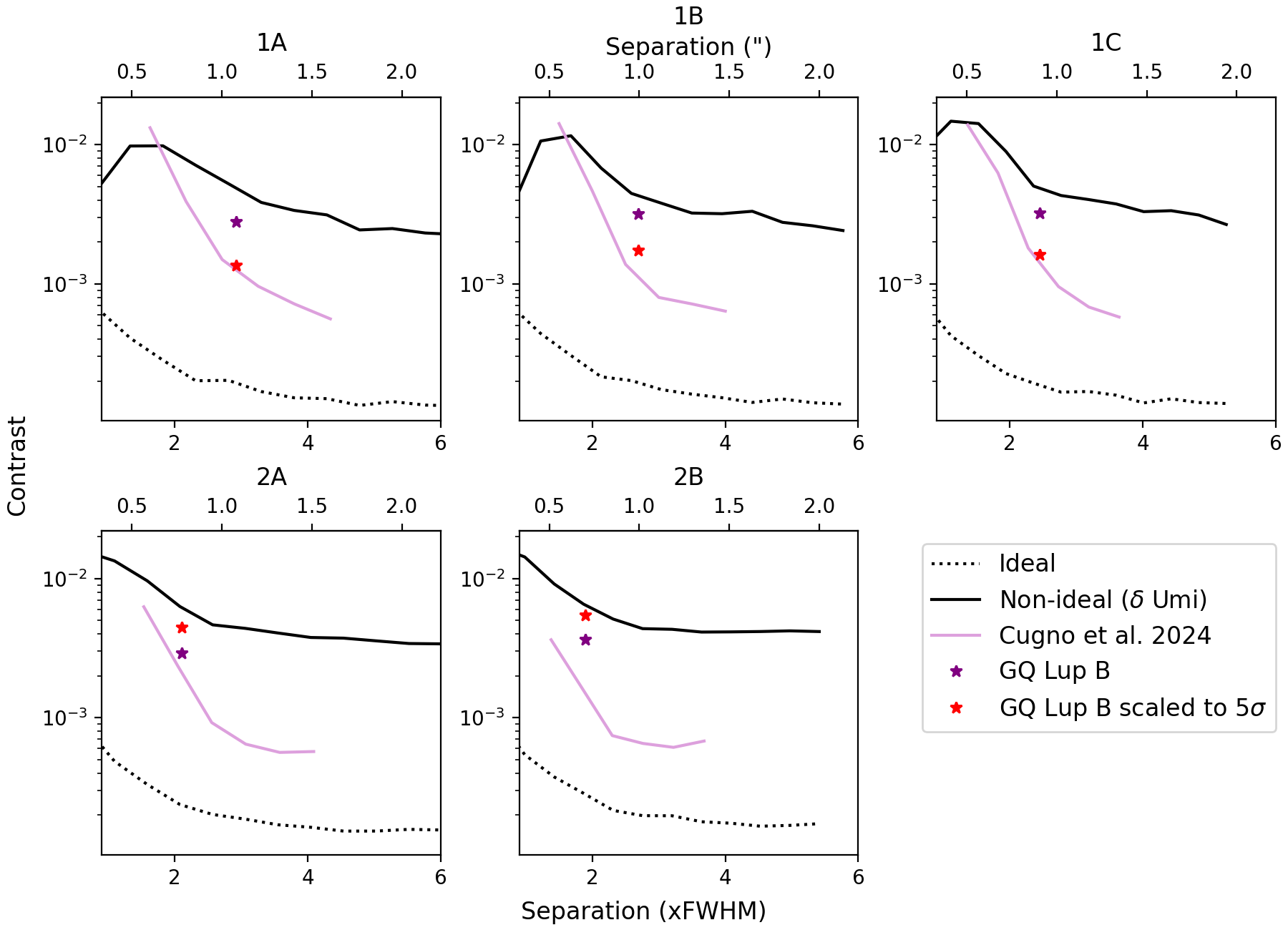}
	\caption{Mean detector-based contrast curves per band, compared to the cube-based results from \citet{ref:24CuPaBa}. The idealised curves are based on a read and photon noise limited case with a total observing time of 37 minutes. The non-ideal case is based on injecting a companion in the subtracted $\delta$~Umi detector images. The red star shows the contrast level for which the current detector-based extracted spectrum of GQ~Lup~B becomes 5$\sigma$ above the noise.}
	\label{fig:contrast_curves}
\end{figure*}

We calculated the contrast curves per band for several cases, which are shown in Fig.~\ref{fig:contrast_curves}. We first take a look at the idealised curve, which is based on a purely read and photon noise limited observation of 37 minutes, near a $\sim$1~Jy star. The idealised noise limit reaches below 10$^{-3}$ levels of contrast even in channel 2, where the background starts to become more significant. Currently, we are not able to reach these idealised curves, though we note that all other observations used as test cases are significantly less deep.

The $\delta$~Umi test data presents a non-idealised case. The star is significantly brighter ($\sim$3~Jy), and the observation much shorter: only 6.5~minutes. In this case, we are clearly not able to go as deep. The contrast limits reach to only $\sim 2 \times 10^{-3}$ at separations of $>$4" at best. We do note that the shape of the curves between the ideal and non-ideal case is very similar, with very shallow increase towards the centre of the PSF. This is possibly due to the undersampling of the PSF, meaning most of the flux is in a single central pixel, rather than more gradually spread over a larger number of pixels, as one might see in a well-sampled cube PSF. This means that large amounts of photon noise only start to appear very close to the central pixel.

Finally, we compare to the GQ~Lup case. The contrast curves derived from the cube-based approach from \citet{ref:24CuPaBa} are shown alongside the resulting contrast of GQ~Lup~B from our detector-based extraction, and the contrast in case we rescale GQ~Lup~B to a 5$\sigma$-detection. In channel 1, the red star mostly overlaps with the cube-based contrast curve, indicating that we almost match. This changes, however, when going to channel 2. In channel 2, we do not reach 5$\sigma$ on the S/N of the spectrum, meaning the achievable contrast is significantly worse compared to the results from  \citet{ref:24CuPaBa}. We omitted the results in band 2C for this reason. However, we already noted that the S/N of our reference star drops in this case, and the reference files are therefore noisier. A similar performance to channel 1 can likely be reached once higher S/N data are available. Furthermore, we expect to be able to get closer to the ideal limit in general when including more data in the PCA, and better cleaning of the reference files based on Papers~I and II.

\section{Discussion}
\label{sec:discussion}

\subsection{Future improvements to the spectral extraction}
As was the case in Papers~I and II, we are limited by the data available to us. By design, the calibration star only provides a sufficient S/N and dynamic range to significantly improve the shorter wavelength ranges of the MRS. Therefore, a first great improvement would be to perform a similar scan on a red object, such that the charge migration correction can be extended to longer wavelengths, and we can derive higher quality fringe flats. Furthermore, the current data do not provide high enough quality PSFs in channels 3 and 4, so the additional grid would also allow us to set up the same spectral extraction framework for the longer wavelengths. Fortunately, the accepted Cycle~5 GO programme 10749 (PI: D. Gasman) was set up to do just that.

An additional factor that we believe would have a large impact is the handling of the pixel weights. The resulting spectral quality is very sensitive to the variances. 
This could be an indication that a better way of estimating the variance needs to be further investigated.

\subsection{High contrast imaging}
\subsubsection{PSF subtraction}
While we show that we can very clearly reveal a faint close companion using both the direct subtraction and PCA approaches, the spectral extraction presented in this work is not yet optimal. We noticed that it appears to be consistently harder to extract a good spectrum from specific dither positions. This is because some of the PSF is more difficult to subtract in specific dither positions, at least using the current setup. The limiting factor for our high-contrast imaging performance is then, perhaps unexpectedly, the achievable quality of the PSF subtraction. Even using our most aggressive PCA approach, the cube images appear extremely clean of the PSF, but the extracted spectrum does not. This means that we opted for the less aggressive method that results in a cleaner spectral extraction, but not the cleanest looking cube.

We based our PCA approach on whichever gave us the most stable and clean spectral extraction, which is currently a purely trial-and-error based process. This is achieved when we only combine the same wavelengths into a PCA PSF model. However, as it stands this is then based on a limited amount of data: only 9 sets, so only 9 components maximum. To illustrate, we compared our results to  \citet{ref:24CuPaBa},  where a larger library is used that includes 17 reference PSFs.  This is one more factor that can bring us closer to the theoretical limit presented in the contrast curves. A more comprehensive set of PSFs would also allow for a more quantitative measure for the optimal number of components to use for PSF subtraction. Once more, additional data would likely be the solution here. We can potentially pull other unresolved sources observed in the nominal dither positions from the archive, process them through our pipeline, and use these in the PCA to improve the quality, but this falls outside the scope of this work. Aside from increasing the number of PSFs in the PCA, additional improvements could be masking the companion, which we experimented with but did not include in the final result. Future experimentation using high-contrast datasets will help improve our methods and allow us to recommend optimal setups.

An additional systematic is the oversubtraction due to the persistence signatures. In order to prevent the PSF subtraction from trying to fit this part of the image, which we know cannot be accurately reproduced by a significantly saturating reference, masking these areas could potentially improve the results. Furthermore, while our test case was the best publicly available dataset for this demonstration, being observed with target acquisition in the correct dither setup, the lack of dedicated background worked against us. Improvements can likely also be achieved when a dataset includes a dedicated background, such that this does not need to be included in the PCA subtraction. Alternative methods to deal with the background could also help. While a PCA approach based on only backgrounds did not yield the desirable background subtraction quality, and was therefore not included further in this work, removing a constant wavelength dependent baseline in the science data might be more effective.

\subsubsection{Caveats of the contrast curves}
We note that the detector-based approximation of a method that is likely better suited to a cube format may not be one-to-one translatable. We attempted to generate contrast curves that were as representative as possible of the curves one might get from a cube-based approach or imager-type data, where a 2D PSF similar to an Airy-disk is surrounded by background. However, the PSF has a significantly different geometry when sliced and dispersed. In our case the fluxes can be distributed differently amongst the slices, meaning the difference between the brightest and second-brightest peaks can be small. Depending on how the PSF is sampled in the $\beta$-dimension, this would likely result in variation in the contrast curve of factor of a few, though not an order of magnitude. Furthermore, the depth of the observation can greatly influence the contrast that can be reached. We did not vary the exposure times or fluxes of the central star, although the variety of test datasets already provides some indication of how these factors influence the contrast curves. Finally, we assumed that the companion is optimally offset on the detector. When the companion overlaps with the central star in the $\alpha$ coordinate, it becomes significantly harder to disentangle the two. Furthermore, the amount of photon noise will almost always be at a maximum, unless the offset between the central star and companion is very large. All of the aforementioned factors mean that many variations within the contrast curves can exist, which we did not explore in this work.

\section{Conclusion}
\label{sec:conclusion}

In this paper, we demonstrated for the first time the feasibility and potential of detector-based optimal spectral extraction and high-contrast imaging applications for MIRI/MRS. While currently the performance of this pipeline exceeds the single object standard pipeline cube extractions by a factor of $>$1.5, at best it matches more involved and specialised approaches of processing cube data. However, more importantly, a well-functioning framework is now in place which can be improved upon further in the future. Additional data can be used to increase the current PSF model S/N and that of point source fringe flats, and expand non-linearity corrections resulting from this series of papers. A part of this will be possible during Cycle~5, when a red target will be observed in a grid as part of PID 10749. Since optimal extraction proves to be highly sensitive to the quality of the reference files, further cleaning of the standard reference files currently in the pipeline will also improve the extracted spectra. Lastly, the pixel variances play an important role in beating down the noise. Better handling of these will also be a valuable upgrade to the current methods. By doing so, we will be able to get closer to the ideal contrast limit, of the order of 10$^{-4}$. Finally, we believe this work has great merit for future integral field spectrometers that use a similar detector setup, such as the Mid-infrared ELT Imager and Spectrograph (METIS) and the ArmazoNes high Dispersion Echelle Spectrograph (ANDES) of ESO's Extremely Large Telescope (ELT).

\section{Data availability}
The code used in this work can be found at \url{https://github.com/DannyG20/MIRI-MRS-Library-Pipeline}, which also includes the methods from Papers~I and II. Future updates and improvements will be directly uploaded there as well.

\begin{acknowledgements}

We thank the referee for their thorough and detailed feedback that greatly improved the clarity of the manuscript.

This work is based on observations made with the NASA/ESA/CSA James Webb Space Telescope. The data were obtained from the Mikulski Archive for Space Telescopes at the Space Telescope Science Institute, which is operated by the Association of Universities for Research in Astronomy, Inc., under NASA contract NAS 5-03127 for JWST. These observations are associated with programmes \#3779, \#1536, \#3886, and \#1640. \\

Parts of the results in this work make use of the colourmaps in the CMasher package. \\

I.A., O.A., and V.C. thank the European Space Agency (ESA) and the Belgian Federal Science Policy Office (BELSPO) for their support in the framework of the PRODEX Programme. \\

P.J.K. acknowledges support from the Research Ireland Pathway programme under Grant Number 21/PATH-S/9360. \\

The work presented is the effort of the entire MIRI team and the enthusiasm within the MIRI partnership is a significant factor in its success. MIRI draws on the scientific and technical expertise of the following organisations: Ames Research Center, USA; Airbus Defence and Space, UK; CEA-Irfu, Saclay, France; Centre Spatial de Liège, Belgium; Consejo Superior de Investigaciones Científicas, Spain; Carl Zeiss Optronics, Germany; Chalmers University of Technology, Sweden; Danish Space Research Institute, Denmark; Dublin Institute for Advanced Studies, Ireland; European Space Agency, Netherlands; ETCA, Belgium; ETH Zurich, Switzerland; Goddard Space Flight Center, USA; Institute d'Astrophysique Spatiale, France; Instituto Nacional de Técnica Aeroespacial, Spain; Institute for Astronomy, Edinburgh, UK; Jet Propulsion Laboratory, USA; Laboratoire d'Astrophysique de Marseille (LAM), France; Leiden University, Netherlands; Lockheed Advanced Technology Center (USA); NOVA Opt-IR group at Dwingeloo, Netherlands; Northrop Grumman, USA; Max-Planck Institut für Astronomie (MPIA), Heidelberg, Germany; Laboratoire d’Etudes Spatiales et d'Instrumentation en Astrophysique (LESIA), France; Paul Scherrer Institut, Switzerland; Raytheon Vision Systems, USA; RUAG Aerospace, Switzerland; Rutherford Appleton Laboratory (RAL Space), UK; Space Telescope Science Institute, USA; Toegepast- Natuurwetenschappelijk Onderzoek (TNO-TPD), Netherlands; UK Astronomy Technology Centre, UK; University College London, UK; University of Amsterdam, Netherlands; University of Arizona, USA; University of Bern, Switzerland; University of Cardiff, UK; University of Cologne, Germany; University of Ghent; University of Groningen, Netherlands; University of Leicester, UK; University of Leuven, Belgium; University of Stockholm, Sweden; Utah State University, USA. A portion of this work was carried out at the Jet Propulsion Laboratory, California Institute of Technology, under a contract with the National Aeronautics and Space Administration.\\

We would like to thank the following National and International Funding Agencies for their support of the MIRI development: NASA; ESA; Belgian Science Policy Office; Centre Nationale D'Etudes Spatiales (CNES); Danish National Space Centre; Deutsches Zentrum für Luft-und Raumfahrt (DLR); Enterprise Ireland; Ministerio de Economía y Competitividad; Netherlands Research School for Astronomy (NOVA); Netherlands Organisation for Scientific Research (NWO); Science and Technology Facilities Council; Swiss Space Office; Swedish National Space Board; UK Space Agency.\\

We take this opportunity to thank the ESA \textit{JWST} Project team and the NASA Goddard ISIM team for their capable technical support in the development of MIRI, its delivery and successful integration.

\end{acknowledgements}

\bibliography{bib}

@ARTICLE{ref:23ArGlLa,
       author = {{Argyriou}, Ioannis and {Glasse}, Alistair and {Law}, David R. and {Labiano}, Alvaro and {{\'A}lvarez-M{\'a}rquez}, Javier and {Patapis}, Polychronis and {Kavanagh}, Patrick J. and {Gasman}, Danny and {Mueller}, Michael and {Larson}, Kirsten and {Vandenbussche}, Bart and {Glauser}, Adrian M. and {Royer}, Pierre and {Dicken}, Daniel and {Harkett}, Jake and {Sargent}, Beth A. and {Engesser}, Michael and {Jones}, Olivia C. and {Kendrew}, Sarah and {Noriega-Crespo}, Alberto and {Brandl}, Bernhard and {Rieke}, George H. and {Wright}, Gillian S. and {Lee}, David and {Wells}, Martyn},
        title = "{JWST MIRI flight performance: The Medium-Resolution Spectrometer}",
      journal = {\aap},
         year = 2023,
        month = jul,
       volume = {675},
          eid = {A111},
        pages = {A111},
          doi = {10.1051/0004-6361/202346489},
archivePrefix = {arXiv},
       eprint = {2303.13469},
 primaryClass = {astro-ph.IM},
       adsurl = {https://ui.adsabs.harvard.edu/abs/2023A&A...675A.111A}
}

@ARTICLE{ref:23WrRiGl,
       author = {{Wright}, Gillian S. and {Rieke}, George H. and {Glasse}, Alistair and {Ressler}, Michael and {Garc{\'\i}a Mar{\'\i}n}, Macarena and {Aguilar}, Jonathan and {Alberts}, Stacey and {{\'A}lvarez-M{\'a}rquez}, Javier and {Argyriou}, Ioannis and {Banks}, Kimberly and {Baudoz}, Pierre and {Boccaletti}, Anthony and {Bouchet}, Patrice and {Bouwman}, Jeroen and {Brandl}, Bernard R. and {Breda}, David and {Bright}, Stacey and {Cale}, Steven and {Colina}, Luis and {Cossou}, Christophe and {Coulais}, Alain and {Cracraft}, Misty and {De Meester}, Wim and {Dicken}, Daniel and {Engesser}, Michael and {Etxaluze}, Mireya and {Fox}, Ori D. and {Friedman}, Scott and {Fu}, Henry and {Gasman}, Danny and {G{\'a}sp{\'a}r}, Andr{\'a}s and {Gastaud}, Ren{\'e} and {Geers}, Vincent and {Glauser}, Adrian Michael and {Gordon}, Karl D. and {Greene}, Thomas and {Greve}, Thomas R. and {Grundy}, Timothy and {G{\"u}del}, Manuel and {Guillard}, Pierre and {Haderlein}, Peter and {Hashimoto}, Ryan and {Henning}, Thomas and {Hines}, Dean and {Holler}, Bryan and {Detre}, {\"O}rs Hunor and {Jahromi}, Amir and {James}, Bryan and {Jones}, Olivia C. and {Justtanont}, Kay and {Kavanagh}, Patrick and {Kendrew}, Sarah and {Klaassen}, Pamela and {Krause}, Oliver and {Labiano}, Alvaro and {Lagage}, Pierre-Olivier and {Lambros}, Scott and {Larson}, Kirsten and {Law}, David and {Lee}, David and {Libralato}, Mattia and {Lorenzo Alverez}, Jose and {Meixner}, Margaret and {Morrison}, Jane and {Mueller}, Migo and {Murray}, Katherine and {Mycroft}, Matthew and {Myers}, Richard and {Nayak}, Omnarayani and {Naylor}, Bret and {Nickson}, Bryony and {Noriega-Crespo}, Alberto and {{\"O}stlin}, G{\"o}ran and {O'Sullivan}, Brian and {Ottens}, Richard and {Patapis}, Polychronis and {Penanen}, Konstantin and {Pietraszkiewicz}, Martin and {Ray}, Tom and {Regan}, Michael and {Roteliuk}, Anthony and {Royer}, Pierre and {Samara-Ratna}, Piyal and {Samuelson}, Bridget and {Sargent}, Beth A. and {Scheithauer}, Silvia and {Schneider}, Analyn and {Schreiber}, J{\"u}rgen and {Shaughnessy}, Bryan and {Sheehan}, Evan and {Shivaei}, Irene and {Sloan}, G.~C. and {Tamas}, Laszlo and {Teague}, Kelly and {Temim}, Tea and {Tikkanen}, Tuomo and {Tustain}, Samuel and {van Dishoeck}, Ewine F. and {Vandenbussche}, Bart and {Weilert}, Mark and {Whitehouse}, Paul and {Wolff}, Schuyler},
        title = "{The Mid-infrared Instrument for JWST and Its In-flight Performance}",
      journal = {\pasp},
         year = 2023,
        month = apr,
       volume = {135},
       number = {1046},
          eid = {048003},
        pages = {048003},
          doi = {10.1088/1538-3873/acbe66},
       adsurl = {https://ui.adsabs.harvard.edu/abs/2023PASP..135d8003W}
}

@ARTICLE{ref:23RiPeMc,
       author = {{Rigby}, Jane and {Perrin}, Marshall and {McElwain}, Michael and {Kimble}, Randy and {Friedman}, Scott and {Lallo}, Matt and {Doyon}, Ren{\'e} and {Feinberg}, Lee and {Ferruit}, Pierre and {Glasse}, Alistair and {Rieke}, Marcia and {Rieke}, George and {Wright}, Gillian and {Willott}, Chris and {Colon}, Knicole and {Milam}, Stefanie and {Neff}, Susan and {Stark}, Christopher and {Valenti}, Jeff and {Abell}, Jim and {Abney}, Faith and {Abul-Huda}, Yasin and {Acton}, D. Scott and {Adams}, Evan and {Adler}, David and {Aguilar}, Jonathan and {Ahmed}, Nasif and {Albert}, Lo{\"\i}c and {Alberts}, Stacey and {Aldridge}, David and {Allen}, Marsha and {Altenburg}, Martin and {{\'A}lvarez-M{\'a}rquez}, Javier and {Alves de Oliveira}, Catarina and {Andersen}, Greg and {Anderson}, Harry and {Anderson}, Sara and {Argyriou}, Ioannis and {Armstrong}, Amber and {Arribas}, Santiago and {Artigau}, Etienne and {Arvai}, Amanda and {Atkinson}, Charles and {Bacon}, Gregory and {Bair}, Thomas and {Banks}, Kimberly and {Barrientes}, Jaclyn and {Barringer}, Bruce and {Bartosik}, Peter and {Bast}, William and {Baudoz}, Pierre and {Beatty}, Thomas and {Bechtold}, Katie and {Beck}, Tracy and {Bergeron}, Eddie and {Bergkoetter}, Matthew and {Bhatawdekar}, Rachana and {Birkmann}, Stephan and {Blazek}, Ronald and {Blome}, Claire and {Boccaletti}, Anthony and {B{\"o}ker}, Torsten and {Boia}, John and {Bonaventura}, Nina and {Bond}, Nicholas and {Bosley}, Kari and {Boucarut}, Ray and {Bourque}, Matthew and {Bouwman}, Jeroen and {Bower}, Gary and {Bowers}, Charles and {Boyer}, Martha and {Bradley}, Larry and {Brady}, Greg and {Braun}, Hannah and {Breda}, David and {Bresnahan}, Pamela and {Bright}, Stacey and {Britt}, Christopher and {Bromenschenkel}, Asa and {Brooks}, Brian and {Brooks}, Keira and {Brown}, Bob and {Brown}, Matthew and {Brown}, Patricia and {Bunker}, Andy and {Burger}, Matthew and {Bushouse}, Howard and {Cale}, Steven and {Cameron}, Alex and {Cameron}, Peter and {Canipe}, Alicia and {Caplinger}, James and {Caputo}, Francis and {Cara}, Mihai and {Carey}, Larkin and {Carniani}, Stefano and {Carrasquilla}, Maria and {Carruthers}, Margaret and {Case}, Michael and {Catherine}, Riggs and {Chance}, Don and {Chapman}, George and {Charlot}, St{\'e}phane and {Charlow}, Brian and {Chayer}, Pierre and {Chen}, Bin and {Cherinka}, Brian and {Chichester}, Sarah and {Chilton}, Zack and {Chonis}, Taylor and {Clampin}, Mark and {Clark}, Charles and {Clark}, Kerry and {Coe}, Dan and {Coleman}, Benee and {Comber}, Brian and {Comeau}, Tom and {Connolly}, Dennis and {Cooper}, James and {Cooper}, Rachel and {Coppock}, Eric and {Correnti}, Matteo and {Cossou}, Christophe and {Coulais}, Alain and {Coyle}, Laura and {Cracraft}, Misty and {Curti}, Mirko and {Cuturic}, Steven and {Davis}, Katherine and {Davis}, Michael and {Dean}, Bruce and {DeLisa}, Amy and {deMeester}, Wim and {Dencheva}, Nadia and {Dencheva}, Nadezhda and {DePasquale}, Joseph and {Deschenes}, Jeremy and {Hunor Detre}, {\"O}rs and {Diaz}, Rosa and {Dicken}, Dan and {DiFelice}, Audrey and {Dillman}, Matthew and {Dixon}, William and {Doggett}, Jesse and {Donaldson}, Tom and {Douglas}, Rob and {DuPrie}, Kimberly and {Dupuis}, Jean and {Durning}, John and {Easmin}, Nilufar and {Eck}, Weston and {Edeani}, Chinwe and {Egami}, Eiichi and {Ehrenwinkler}, Ralf and {Eisenhamer}, Jonathan and {Eisenhower}, Michael and {Elie}, Michelle and {Elliott}, James and {Elliott}, Kyle and {Ellis}, Tracy and {Engesser}, Michael and {Espinoza}, Nestor and {Etienne}, Odessa and {Etxaluze}, Mireya and {Falini}, Patrick and {Feeney}, Matthew and {Ferry}, Malcolm and {Filippazzo}, Joseph and {Fincham}, Brian and {Fix}, Mees and {Flagey}, Nicolas and {Florian}, Michael and {Flynn}, Jim and {Fontanella}, Erin and {Ford}, Terrance and {Forshay}, Peter and {Fox}, Ori and {Franz}, David and {Fu}, Henry and {Fullerton}, Alexander and {Galkin}, Sergey and {Galyer}, Anthony and {Garc{\'\i}a Mar{\'\i}n}, Macarena and {Gardner}, Jonathan P. and {Gardner}, Lisa and {Garland}, Dennis and {Garrett}, Bruce and {Gasman}, Danny and {Gaspar}, Andras and {Gaudreau}, Daniel and {Gauthier}, Peter and {Geers}, Vincent and {Geithner}, Paul and {Gennaro}, Mario and {Giardino}, Giovanna and {Girard}, Julien and {Giuliano}, Mark and {Glassmire}, Kirk and {Glauser}, Adrian and {Glazer}, Stuart and {Godfrey}, John and {Golimowski}, David and {Gollnitz}, David and {Gong}, Fan and {Gonzaga}, Shireen and {Gordon}, Michael and {Gordon}, Karl and {Goudfrooij}, Paul and {Greene}, Thomas and {Greenhouse}, Matthew and {Grimaldi}, Stefano and {Groebner}, Andrew and {Grundy}, Timothy and {Guillard}, Pierre and {Gutman}, Irvin and {Ha}, Kong Q. and {Haderlein}, Peter and {Hagedorn}, Andria and {Hainline}, Kevin and {Haley}, Craig and {Hami}, Maryam and {Hamilton}, Forrest and {Hammel}, Heidi and {Hansen}, Carl and {Harkins}, Tom and {Harr}, Michael and {Hart}, Jessica and {Hart}, Quyen and {Hartig}, George and {Hashimoto}, Ryan and {Haskins}, Sujee and {Hathaway}, William and {Havey}, Keith and {Hayden}, Brian and {Hecht}, Karen and {Heller-Boyer}, Chris and {Henriques}, Caroline and {Henry}, Alaina and {Hermann}, Karl and {Hernandez}, Scarlin and {Hesman}, Brigette and {Hicks}, Brian and {Hilbert}, Bryan and {Hines}, Dean and {Hoffman}, Melissa and {Holfeltz}, Sherie and {Holler}, Bryan J. and {Hoppa}, Jennifer and {Hott}, Kyle and {Howard}, Joseph M. and {Howard}, Rick and {Hunter}, Alexander and {Hunter}, David and {Hurst}, Brendan and {Husemann}, Bernd and {Hustak}, Leah and {Ilinca Ignat}, Luminita and {Illingworth}, Garth and {Irish}, Sandra and {Jackson}, Wallace and {Jahromi}, Amir and {Jakobsen}, Peter and {James}, LeAndrea and {James}, Bryan and {Januszewski}, William and {Jenkins}, Ann and {Jirdeh}, Hussein and {Johnson}, Phillip and {Johnson}, Timothy and {Jones}, Vicki and {Jones}, Ron and {Jones}, Danny and {Jones}, Olivia and {Jordan}, Ian and {Jordan}, Margaret and {Jurczyk}, Sarah and {Jurling}, Alden and {Kaleida}, Catherine and {Kalmanson}, Phillip and {Kammerer}, Jens and {Kang}, Huijo and {Kao}, Shaw-Hong and {Karakla}, Diane and {Kavanagh}, Patrick and {Kelly}, Doug and {Kendrew}, Sarah and {Kennedy}, Herbert and {Kenny}, Deborah and {Keski-kuha}, Ritva and {Keyes}, Charles and {Kidwell}, Richard and {Kinzel}, Wayne and {Kirk}, Jeff and {Kirkpatrick}, Mark and {Kirshenblat}, Danielle and {Klaassen}, Pamela and {Knapp}, Bryan and {Knight}, J. Scott and {Knollenberg}, Perry and {Koehler}, Robert and {Koekemoer}, Anton and {Kovacs}, Aiden and {Kulp}, Trey and {Kumari}, Nimisha and {Kyprianou}, Mark and {La Massa}, Stephanie and {Labador}, Aurora and {Labiano}, Alvaro and {Lagage}, Pierre-Olivier and {Lajoie}, Charles-Philippe and {Lallo}, Matthew and {Lam}, May and {Lamb}, Tracy and {Lambros}, Scott and {Lampenfield}, Richard and {Langston}, James and {Larson}, Kirsten and {Law}, David and {Lawrence}, Jon and {Lee}, David and {Leisenring}, Jarron and {Lepo}, Kelly and {Leveille}, Michael and {Levenson}, Nancy and {Levine}, Marie and {Levy}, Zena and {Lewis}, Dan and {Lewis}, Hannah and {Libralato}, Mattia and {Lightsey}, Paul and {Link}, Miranda and {Liu}, Lily and {Lo}, Amy and {Lockwood}, Alexandra and {Logue}, Ryan and {Long}, Chris and {Long}, Douglas and {Loomis}, Charles and {Lopez-Caniego}, Marcos and {Lorenzo Alvarez}, Jose and {Love-Pruitt}, Jennifer and {Lucy}, Adrian and {Luetzgendorf}, Nora and {Maghami}, Peiman and {Maiolino}, Roberto and {Major}, Melissa and {Malla}, Sunita and {Malumuth}, Eliot and {Manjavacas}, Elena and {Mannfolk}, Crystal and {Marrione}, Amanda and {Marston}, Anthony and {Martel}, Andr{\'e} and {Maschmann}, Marc and {Masci}, Gregory and {Masciarelli}, Michaela and {Maszkiewicz}, Michael and {Mather}, John and {McKenzie}, Kenny and {McLean}, Brian and {McMaster}, Matthew and {Melbourne}, Katie and {Mel{\'e}ndez}, Marcio and {Menzel}, Michael and {Merz}, Kaiya and {Meyett}, Michele and {Meza}, Luis and {Miskey}, Cherie and {Misselt}, Karl and {Moller}, Christopher and {Morrison}, Jane and {Morse}, Ernie and {Moseley}, Harvey and {Mosier}, Gary and {Mountain}, Matt and {Mueckay}, Julio and {Mueller}, Michael and {Mullally}, Susan and {Murphy}, Jess and {Murray}, Katherine and {Murray}, Claire and {Mustelier}, David and {Muzerolle}, James and {Mycroft}, Matthew and {Myers}, Richard and {Myrick}, Kaila and {Nanavati}, Shashvat and {Nance}, Elizabeth and {Nayak}, Omnarayani and {Naylor}, Bret and {Nelan}, Edmund and {Nickson}, Bryony and {Nielson}, Alethea and {Nieto-Santisteban}, Maria and {Nikolov}, Nikolay and {Noriega-Crespo}, Alberto and {O'Shaughnessy}, Brian and {O'Sullivan}, Brian and {Ochs}, William and {Ogle}, Patrick and {Oleszczuk}, Brenda and {Olmsted}, Joseph and {Osborne}, Shannon and {Ottens}, Richard and {Owens}, Beverly and {Pacifici}, Camilla and {Pagan}, Alyssa and {Page}, James and {Park}, Sang and {Parrish}, Keith and {Patapis}, Polychronis and {Paul}, Lee and {Pauly}, Tyler and {Pavlovsky}, Cheryl and {Pedder}, Andrew and {Peek}, Matthew and {Pena-Guerrero}, Maria and {Penanen}, Konstantin and {Perez}, Yesenia and {Perna}, Michele and {Perriello}, Beth and {Phillips}, Kevin and {Pietraszkiewicz}, Martin and {Pinaud}, Jean-Paul and {Pirzkal}, Norbert and {Pitman}, Joseph and {Piwowar}, Aidan and {Platais}, Vera and {Player}, Danielle and {Plesha}, Rachel and {Pollizi}, Joe and {Polster}, Ethan and {Pontoppidan}, Klaus and {Porterfield}, Blair and {Proffitt}, Charles and {Pueyo}, Laurent and {Pulliam}, Christine and {Quirt}, Brian and {Quispe Neira}, Irma and {Ramos Alarcon}, Rafael and {Ramsay}, Leah and {Rapp}, Greg and {Rapp}, Robert and {Rauscher}, Bernard and {Ravindranath}, Swara and {Rawle}, Timothy and {Regan}, Michael and {Reichard}, Timothy A. and {Reis}, Carl and {Ressler}, Michael E. and {Rest}, Armin and {Reynolds}, Paul and {Rhue}, Timothy and {Richon}, Karen and {Rickman}, Emily and {Ridgaway}, Michael and {Ritchie}, Christine and {Rix}, Hans-Walter and {Robberto}, Massimo and {Robinson}, Gregory and {Robinson}, Michael and {Robinson}, Orion and {Rock}, Frank and {Rodriguez}, David and {Rodriguez Del Pino}, Bruno and {Roellig}, Thomas and {Rohrbach}, Scott and {Roman}, Anthony and {Romelfanger}, Fred and {Rose}, Perry and {Roteliuk}, Anthony and {Roth}, Marc and {Rothwell}, Braden and {Rowlands}, Neil and {Roy}, Arpita and {Royer}, Pierre and {Royle}, Patricia and {Rui}, Chunlei and {Rumler}, Peter and {Runnels}, Joel and {Russ}, Melissa and {Rustamkulov}, Zafar and {Ryden}, Grant and {Ryer}, Holly and {Sabata}, Modhumita and {Sabatke}, Derek and {Sabbi}, Elena and {Samuelson}, Bridget and {Sapp}, Benjamin and {Sappington}, Bradley and {Sargent}, B. and {Sauer}, Arne and {Scheithauer}, Silvia and {Schlawin}, Everett and {Schlitz}, Joseph and {Schmitz}, Tyler and {Schneider}, Analyn and {Schreiber}, J{\"u}rgen and {Schulze}, Vonessa and {Schwab}, Ryan and {Scott}, John and {Sembach}, Kenneth and {Shanahan}, Clare and {Shaughnessy}, Bryan and {Shaw}, Richard and {Shawger}, Nanci and {Shay}, Christopher and {Sheehan}, Evan and {Shen}, Sharon and {Sherman}, Allan and {Shiao}, Bernard and {Shih}, Hsin-Yi and {Shivaei}, Irene and {Sienkiewicz}, Matthew and {Sing}, David and {Sirianni}, Marco and {Sivaramakrishnan}, Anand and {Skipper}, Joy and {Sloan}, G.~C. and {Slocum}, Christine and {Slowinski}, Steven and {Smith}, Erin and {Smith}, Eric and {Smith}, Denise and {Smith}, Corbett and {Snyder}, Gregory and {Soh}, Warren and {Sohn}, Sangmo Tony and {Soto}, Christian and {Spencer}, Richard and {Stallcup}, Scott and {Stansberry}, John and {Starr}, Carl and {Starr}, Elysia and {Stewart}, Alphonso and {Stiavelli}, Massimo and {Straughn}, Amber and {Strickland}, David and {Stys}, Jeff and {Summers}, Francis and {Sun}, Fengwu and {Sunnquist}, Ben and {Swade}, Daryl and {Swam}, Michael and {Swaters}, Robert and {Swoish}, Robby and {Taylor}, Joanna M. and {Taylor}, Rolanda and {Te Plate}, Maurice and {Tea}, Mason and {Teague}, Kelly and {Telfer}, Randal and {Temim}, Tea and {Thatte}, Deepashri and {Thompson}, Christopher and {Thompson}, Linda and {Thomson}, Shaun and {Tikkanen}, Tuomo and {Tippet}, William and {Todd}, Connor and {Toolan}, Sharon and {Tran}, Hien and {Trejo}, Edwin and {Truong}, Justin and {Tsukamoto}, Chris and {Tustain}, Samuel and {Tyra}, Harrison and {Ubeda}, Leonardo and {Underwood}, Kelli and {Uzzo}, Michael and {Van Campen}, Julie and {Vandal}, Thomas and {Vandenbussche}, Bart and {Vila}, Bego{\~n}a and {Volk}, Kevin and {Wahlgren}, Glenn and {Waldman}, Mark and {Walker}, Chanda and {Wander}, Michel and {Warfield}, Christine and {Warner}, Gerald and {Wasiak}, Matthew and {Watkins}, Mitchell and {Weaver}, Andrew and {Weilert}, Mark and {Weiser}, Nick and {Weiss}, Ben and {Weissman}, Sarah and {Welty}, Alan and {West}, Garrett and {Wheate}, Lauren and {Wheatley}, Elizabeth and {Wheeler}, Thomas and {White}, Rick and {Whiteaker}, Kevin and {Whitehouse}, Paul and {Whiteleather}, Jennifer and {Whitman}, William and {Williams}, Christina and {Willmer}, Christopher and {Willoughby}, Scott and {Wilson}, Andrew and {Wirth}, Gregory and {Wislowski}, Emily and {Wolf}, Erin and {Wolfe}, David and {Wolff}, Schuyler and {Workman}, Bill and {Wright}, Ray and {Wu}, Carl and {Wu}, Rai and {Wymer}, Kristen and {Yates}, Kayla and {Yeager}, Christopher and {Yeates}, Jared and {Yerger}, Ethan and {Yoon}, Jinmi and {Young}, Alice and {Yu}, Susan and {Zak}, Dean and {Zeidler}, Peter and {Zhou}, Julia and {Zielinski}, Thomas and {Zincke}, Cristian and {Zonak}, Stephanie},
        title = "{The Science Performance of JWST as Characterized in Commissioning}",
      journal = {\pasp},
         year = 2023,
        month = apr,
       volume = {135},
       number = {1046},
          eid = {048001},
        pages = {048001},
          doi = {10.1088/1538-3873/acb293},
archivePrefix = {arXiv},
       eprint = {2207.05632},
 primaryClass = {astro-ph.IM},
       adsurl = {https://ui.adsabs.harvard.edu/abs/2023PASP..135d8001R}
}

@ARTICLE{ref:20ArWeGl,
       author = {{Argyriou}, Ioannis and {Wells}, Martyn and {Glasse}, Alistair and {Lee}, David and {Royer}, Pierre and {Vandenbussche}, Bart and {Malumuth}, Eliot and {Glauser}, Adrian and {Kavanagh}, Patrick J. and {Labiano}, Alvaro and {Lahuis}, Fred and {Mueller}, Michael and {Patapis}, Polychronis},
        title = "{The nature of point source fringes in mid-infrared spectra acquired with the James Webb Space Telescope}",
      journal = {\aap},
         year = 2020,
        month = sep,
       volume = {641},
          eid = {A150},
        pages = {A150},
          doi = {10.1051/0004-6361/202037535},
archivePrefix = {arXiv},
       eprint = {2007.16143},
 primaryClass = {astro-ph.IM},
       adsurl = {https://ui.adsabs.harvard.edu/abs/2020A&A...641A.150A}
}

@ARTICLE{paperI,
       author = {{Gasman}, Danny and {Argyriou}, Ioannis and {Morrison}, Jane E. and {Law}, David R. and {Glasse}, Alistair and {Gordon}, Karl D. and {Kavanagh}, Patrick J. and {Lage}, Craig and {Patapis}, Polychronis and {Sloan}, Gregory C.},
        title = "{The MIRI/MRS Library. I. Empirically correcting detector charge migration in unresolved sources}",
      journal = {\aap},
         year = 2024,
        month = aug,
       volume = {688},
          eid = {A226},
        pages = {A226},
          doi = {10.1051/0004-6361/202450241},
archivePrefix = {arXiv},
       eprint = {2406.10835},
 primaryClass = {astro-ph.IM},
       adsurl = {https://ui.adsabs.harvard.edu/abs/2024A&A...688A.226G}
}

@ARTICLE{ref:25LaArGo,
       author = {{Law}, David R. and {Argyriou}, Ioannis and {Gordon}, Karl D. and {Sloan}, G.~C. and {Gasman}, Danny and {Glasse}, Alistair and {Larson}, Kirsten and {Fletcher}, Leigh N. and {Labiano}, Alvaro and {Noriega-Crespo}, Alberto},
        title = "{The James Webb Space Telescope Absolute Flux Calibration. III. Mid-infrared Instrument Medium Resolution Integral Field Unit Spectrometer}",
      journal = {\aj},
         year = 2025,
        month = feb,
       volume = {169},
       number = {2},
          eid = {67},
        pages = {67},
          doi = {10.3847/1538-3881/ad9685},
archivePrefix = {arXiv},
       eprint = {2409.15435},
 primaryClass = {astro-ph.IM},
       adsurl = {https://ui.adsabs.harvard.edu/abs/2025AJ....169...67L}
}

@ARTICLE{ref:23LaMoAr,
       author = {{Law}, David R. and {E. Morrison}, Jane and {Argyriou}, Ioannis and {Patapis}, Polychronis and {{\'A}lvarez-M{\'a}rquez}, J. and {Labiano}, Alvaro and {Vandenbussche}, Bart},
        title = "{A 3D Drizzle Algorithm for JWST and Practical Application to the MIRI Medium Resolution Spectrometer}",
      journal = {\aj},
         year = 2023,
        month = aug,
       volume = {166},
       number = {2},
          eid = {45},
        pages = {45},
          doi = {10.3847/1538-3881/acdddc},
archivePrefix = {arXiv},
       eprint = {2306.05520},
 primaryClass = {astro-ph.IM},
       adsurl = {https://ui.adsabs.harvard.edu/abs/2023AJ....166...45L}
}

@ARTICLE{ref:23ArLaRi,
       author = {{Argyriou}, Ioannis and {Lage}, Craig and {Rieke}, George H. and {Gasman}, Danny and {Bouwman}, Jeroen and {Morrison}, Jane and {Libralato}, Mattia and {Dicken}, Daniel and {Brandl}, Bernhard R. and {{\'A}lvarez-M{\'a}rquez}, Javier and {Labiano}, Alvaro and {Regan}, Michael and {Ressler}, Michael E.},
        title = "{The brighter-fatter effect in the JWST MIRI Si:As IBC detectors. I. Observations, impact on science, and modeling}",
      journal = {\aap},
         year = 2023,
        month = dec,
       volume = {680},
          eid = {A96},
        pages = {A96},
          doi = {10.1051/0004-6361/202346490},
archivePrefix = {arXiv},
       eprint = {2303.13517},
 primaryClass = {astro-ph.IM},
       adsurl = {https://ui.adsabs.harvard.edu/abs/2023A&A...680A..96A}
}

@ARTICLE{paperII,
       author = {{Gasman}, Danny and {Argyriou}, Ioannis and {Law}, David R. and {Glasse}, Alistair and {Gordon}, Karl D. and {Kavanagh}, Patrick J. and {Morrison}, Jane E. and {Patapis}, Polychronis and {Sloan}, Gregory C.},
        title = "{The MIRI/MRS Library: II. Pointing-based defringing of unresolved sources}",
      journal = {\aap},
         year = 2025,
        month = may,
       volume = {697},
          eid = {A58},
        pages = {A58},
          doi = {10.1051/0004-6361/202554055},
       adsurl = {https://ui.adsabs.harvard.edu/abs/2025A&A...697A..58G}
}

@ARTICLE{ref:86Ho,
       author = {{Horne}, K.},
        title = "{An optimal extraction algorithm for CCD spectroscopy.}",
      journal = {\pasp},
         year = 1986,
        month = jun,
       volume = {98},
        pages = {609-617},
          doi = {10.1086/131801},
       adsurl = {https://ui.adsabs.harvard.edu/abs/1986PASP...98..609H}
}

@ARTICLE{ref:89Ma,
       author = {{Marsh}, T.~R.},
        title = "{The Extraction of Highly Distorted Spectra}",
      journal = {\pasp},
         year = 1989,
        month = nov,
       volume = {101},
        pages = {1032},
          doi = {10.1086/132570},
       adsurl = {https://ui.adsabs.harvard.edu/abs/1989PASP..101.1032M}
}

@INPROCEEDINGS{ref:92Va,
       author = {{Valdes}, F.},
        title = "{A Quantitative Study of Optimal Extraction}",
    booktitle = {Astronomical Data Analysis Software and Systems I},
         year = 1992,
       editor = {{Worrall}, Diana M. and {Biemesderfer}, Chris and {Barnes}, Jeannette},
       series = {Astronomical Society of the Pacific Conference Series},
       volume = {25},
        month = jan,
        pages = {398},
       adsurl = {https://ui.adsabs.harvard.edu/abs/1992ASPC...25..398V}
}

@ARTICLE{ref:24RuPeHo,
       author = {{Ruffio}, Jean-Baptiste and {Perrin}, Marshall D. and {Hoch}, Kielan K.~W. and {Kammerer}, Jens and {Konopacky}, Quinn M. and {Pueyo}, Laurent and {Madurowicz}, Alex and {Rickman}, Emily and {Theissen}, Christopher A. and {Agrawal}, Shubh and {Greenbaum}, Alexandra Z. and {Miles}, Brittany E. and {Barman}, Travis S. and {Balmer}, William O. and {Llop-Sayson}, Jorge and {Girard}, Julien H. and {Rebollido}, Isabel and {Soummer}, R{\'e}mi and {Allen}, Natalie H. and {Anderson}, Jay and {Beichman}, Charles A. and {Bellini}, Andrea and {Bryden}, Geoffrey and {Espinoza}, N{\'e}stor and {Glidden}, Ana and {Huang}, Jingcheng and {Lewis}, Nikole K. and {Libralato}, Mattia and {Louie}, Dana R. and {Sohn}, Sangmo Tony and {Seager}, Sara and {van der Marel}, Roeland P. and {Wakeford}, Hannah R. and {Watkins}, Laura L. and {Ygouf}, Marie and {Mountain}, C. Matt},
        title = "{JWST-TST High Contrast: Achieving Direct Spectroscopy of Faint Substellar Companions Next to Bright Stars with the NIRSpec Integral Field Unit}",
      journal = {\aj},
         year = 2024,
        month = aug,
       volume = {168},
       number = {2},
          eid = {73},
        pages = {73},
          doi = {10.3847/1538-3881/ad5281},
archivePrefix = {arXiv},
       eprint = {2310.09902},
 primaryClass = {astro-ph.EP},
       adsurl = {https://ui.adsabs.harvard.edu/abs/2024AJ....168...73R}
}

@ARTICLE{ref:07JeBlSi,
       author = {{Jee}, M.~J. and {Blakeslee}, J.~P. and {Sirianni}, M. and {Martel}, A.~R. and {White}, R.~L. and {Ford}, H.~C.},
        title = "{Principal Component Analysis of the Time- and Position-dependent Point-Spread Function of the Advanced Camera for Surveys}",
      journal = {\pasp},
         year = 2007,
        month = dec,
       volume = {119},
       number = {862},
        pages = {1403-1419},
          doi = {10.1086/524849},
archivePrefix = {arXiv},
       eprint = {0710.5560},
 primaryClass = {astro-ph},
       adsurl = {https://ui.adsabs.harvard.edu/abs/2007PASP..119.1403J}
}

@ARTICLE{ref:26RuXuCh,
       author = {{Ruffio}, Jean-Baptiste and {Xuan}, Jerry W. and {Chachan}, Yayaati and {Kesseli}, Aurora and {Lee}, Eve J. and {Beichman}, Charles and {Hodapp}, Klaus and {Balmer}, William O. and {Konopacky}, Quinn and {Perrin}, Marshall D. and {Mawet}, Dimitri and {Knutson}, Heather A. and {Bryden}, Geoffrey and {Greene}, Thomas P. and {Johnstone}, Doug and {Leisenring}, Jarron and {Meyer}, Michael and {Ygouf}, Marie},
        title = "{Jupiter-like uniform metal enrichment in a system of multiple giant exoplanets}",
      journal = {Nature Astronomy},
         year = 2026,
        month = apr,
       volume = {10},
        pages = {511-521},
          doi = {10.1038/s41550-026-02783-z},
archivePrefix = {arXiv},
       eprint = {2601.08227},
 primaryClass = {astro-ph.EP},
       adsurl = {https://ui.adsabs.harvard.edu/abs/2026NatAs..10..511R}
}

@ARTICLE{ref:24CuPaBa,
       author = {{Cugno}, Gabriele and {Patapis}, Polychronis and {Banzatti}, Andrea and {Meyer}, Michael and {Dannert}, Felix A. and {Stolker}, Tomas and {MacDonald}, Ryan J. and {Pontoppidan}, Klaus M.},
        title = "{Mid-infrared Spectrum of the Disk around the Forming Companion GQ Lup B Revealed by JWST/MIRI}",
      journal = {\apjl},
         year = 2024,
        month = may,
       volume = {966},
       number = {1},
          eid = {L21},
        pages = {L21},
          doi = {10.3847/2041-8213/ad3cbc},
archivePrefix = {arXiv},
       eprint = {2404.07086},
 primaryClass = {astro-ph.EP},
       adsurl = {https://ui.adsabs.harvard.edu/abs/2024ApJ...966L..21C}
}

@ARTICLE{ref:12SoLaJa,
       author = {{Soummer}, R{\'e}mi and {Pueyo}, Laurent and {Larkin}, James},
        title = "{Detection and Characterization of Exoplanets and Disks Using Projections on Karhunen-Lo{\`e}ve Eigenimages}",
      journal = {\apjl},
         year = 2012,
        month = aug,
       volume = {755},
       number = {2},
          eid = {L28},
        pages = {L28},
          doi = {10.1088/2041-8205/755/2/L28},
archivePrefix = {arXiv},
       eprint = {1207.4197},
 primaryClass = {astro-ph.IM},
       adsurl = {https://ui.adsabs.harvard.edu/abs/2012ApJ...755L..28S}
}

@ARTICLE{ref:25MaRuMa,
       author = {{Madurowicz}, Alexander and {Ruffio}, Jean-Baptiste and {Macintosh}, Bruce and {Perrin}, Marshall and {Konopacky}, Quinn M. and {Baburaj}, Aneesh and {Hoch}, Kielan},
        title = "{Direct Spectroscopy of 51 Eridani b with JWST NIRSpec}",
      journal = {\aj},
         year = 2025,
        month = dec,
       volume = {170},
       number = {6},
          eid = {326},
        pages = {326},
          doi = {10.3847/1538-3881/ae1028},
archivePrefix = {arXiv},
       eprint = {2510.08327},
 primaryClass = {astro-ph.EP},
       adsurl = {https://ui.adsabs.harvard.edu/abs/2025AJ....170..326M}
}

@ARTICLE{ref:05NeGuWu,
       author = {{Neuh{\"a}user}, R. and {Guenther}, E.~W. and {Wuchterl}, G. and {Mugrauer}, M. and {Bedalov}, A. and {Hauschildt}, P.~H.},
        title = "{Evidence for a co-moving sub-stellar companion of GQ Lup}",
      journal = {\aap},
         year = 2005,
        month = may,
       volume = {435},
       number = {1},
        pages = {L13-L16},
          doi = {10.1051/0004-6361:200500104},
archivePrefix = {arXiv},
       eprint = {astro-ph/0503691},
 primaryClass = {astro-ph},
       adsurl = {https://ui.adsabs.harvard.edu/abs/2005A&A...435L..13N}
}

@ARTICLE{ref:cmasher,
       author = {{van der Velden}, Ellert},
        title = "{CMasher: Scientific colormaps for making accessible, informative and 'cmashing' plots}",
      journal = {The Journal of Open Source Software},
         year = 2020,
        month = feb,
       volume = {5},
       number = {46},
          eid = {2004},
        pages = {2004},
          doi = {10.21105/joss.02004},
archivePrefix = {arXiv},
       eprint = {2003.01069},
 primaryClass = {eess.IV},
       adsurl = {https://ui.adsabs.harvard.edu/abs/2020JOSS....5.2004V}
}
\bibliographystyle{aa}

\begin{appendix}
\onecolumn
\section{Additional PSF subtraction plots}
\label{app:psf_sub}

\begin{figure*}[h!]
	\centering
	\includegraphics[width=0.75\textwidth]{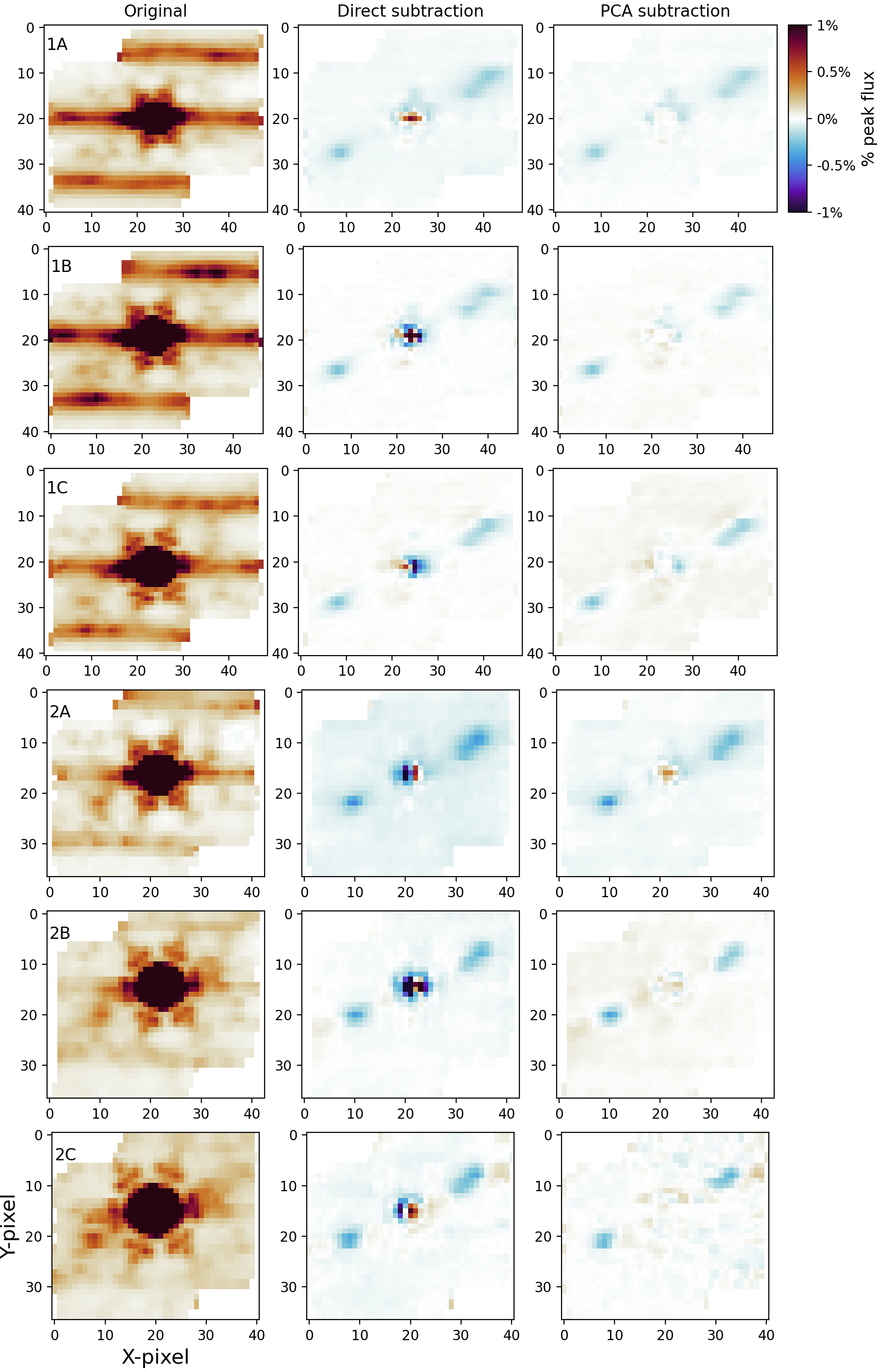}
	\caption{Performance of the different detector-based PSF subtraction methods per band, applied to a single calibration star ($\delta$~Umi).  The first column presents the original data, the second column the residuals of direct subtraction using the detector-based models, and the third column the subtraction using PCA. We note that all cube panels present the median-collapsed cubes, in percent of the peak of the original data, whereas the detector-based results is a percent of the maximum flux in the image. Colourmap from \citet{ref:cmasher}.}
	\label{fig:star_subtr_full}
\end{figure*}

\begin{figure*}[h!]
	\centering
	\includegraphics[width=0.75\textwidth]{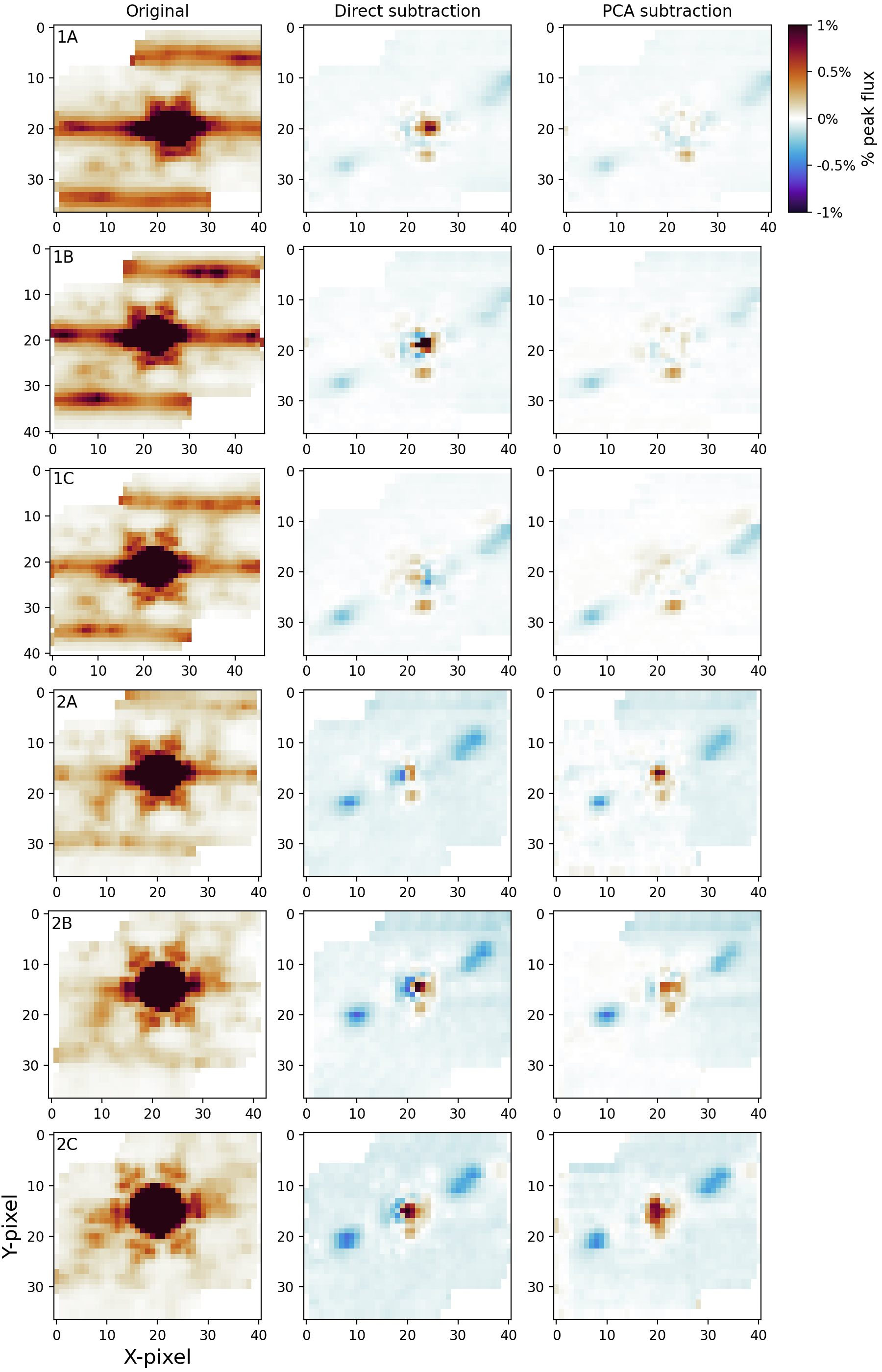}
	\caption{Same as Fig.~\ref{fig:star_subtr_full}, applied to an object known to host a companion (GQ~Lup A\&B) at a separation of $\sim0\farcs7$. }
	\label{fig:gqlup_subtr_full}
\end{figure*}

The cube images of the PSF subtraction per individual band can be found here. $\delta$~Umi is shown in Fig.~\ref{fig:star_subtr_full}, and GQ~Lup in Fig.~\ref{fig:gqlup_subtr_full}.
\FloatBarrier

\end{appendix}

\end{document}